\documentclass[aps,prx,groupedaddress,eqsecnum,preprintnumbers,superscriptaddress,dblfloatfix,nofootinbib,twocolumn]{revtex4-2}
\usepackage[utf8]{inputenc}
\usepackage{hyperref}
\usepackage{graphicx}
\usepackage{amssymb,amsmath}
\usepackage{epstopdf}
\usepackage{enumitem} 
\usepackage{ulem} 
\usepackage[table,dvipsnames]{xcolor}
\usepackage{bm}
\usepackage{tabularx}
\usepackage{booktabs}
\DeclareMathAlphabet{\mathpzc}{OT1}{pzc}{m}{it}

\newcommand{\dd}{\text{d}}

\begin{document}

\title{Gravitational wave constraints on the Paneitz operator}

\author{\textsc{Robin Valtin}}
    \affiliation{ Institute for Theoretical Physics, Leibniz University Hannover,\\ Appelstraße 2, 30167 Hannover, Germany}
\author{\textsc{Alexander Ganz}}
   \email{{alexander.ganz}@{itp.uni-hannover.de}}
    \affiliation{ Institute for Theoretical Physics, Leibniz University Hannover,\\ Appelstraße 2, 30167 Hannover, Germany}
\author{\textsc{Guillem Dom\`enech }}
    \email{{guillem.domenech}@{itp.uni-hannover.de}}
    \affiliation{ Institute for Theoretical Physics, Leibniz University Hannover,\\ Appelstraße 2, 30167 Hannover, Germany}
        \affiliation{ Max Planck Institute for Gravitational Physics, Albert Einstein Institute, 30167 Hannover, Germany}

\begin{abstract}
The Paneitz operator is a dimension-4 conformally invariant fourth-order differential operator that has recently attracted attention for possible cancellations of the vacuum energy. We show that, in four dimensions, the Paneitz operator acting on a scalar field falls within the class of extended mimetic gravity theories. Thus, it exhibits the usual instabilities of mimetic gravity. Assuming such instabilities are cured by higher derivative terms, we derive constraints on the Paneitz operator from a modified propagation speed of gravitational waves, after including the Einstein-Hilbert action in the mimetic gravity formulation.
\end{abstract}

\maketitle

\section{Introduction}

Symmetries have been a guiding principle in understanding and constructing new fundamental theories in physics, in particular in the Standard Model of particle physics. Scale symmetry could be one symmetry of quantum gravity that is broken at low energies \cite{Wetterich:2019qzx,Wetterich:2020cxq}. In fact, the breaking of scale symmetry has been studied in the context of the hierarchy problem \cite{Wetterich:1983bi,Hempfling:1996ht,Meissner:2006zh,Foot:2007as,Aoki:2012xs,Wetterich:2016uxm}, the cosmological constant \cite{Wetterich:1987fm,Rabinovici:1987tf}, and cosmic inflation \cite{Rubio:2017gty,Rubio:2018ogq}.
Furthermore, observations of the Cosmic Microwave Background (CMB) measured an almost scale-invariant spectrum of primordial fluctuations \cite{Planck:2018vyg,Planck:2018jri}, explained by the almost de Sitter expansion during cosmic inflation.

For similar reasons, conformal (or Weyl) symmetry, that is, a local scale symmetry \cite{Nakayama:2013is}, may be an appealing feature of theories of gravity \cite{Mannheim:2011ds,Lucat:2016eze,Rachwal:2018gwu}. Additionally, conformal symmetries play a key role in the AdS/CFT correspondence \cite{Witten_1998,Maldacena_1999,Fitzpatrick_2013}. Interestingly, when it involves an additional scalar field, conformal symmetry may give rise to mimetic dark matter \cite{Chamseddine:2013kea}. This is the case of a conformal invariant auxiliary metric \cite{Chamseddine:2013kea,Golovnev:2013jxa,Barvinsky:2013mea} (also interpreted as a singular metric transformation \cite{Deruelle:2014zza,Arroja:2015wpa,Jirousek:2022rym,Golovnev:2022jts,Jirousek:2022jhh}) and a conformal invariant action including higher derivatives of the scalar field \cite{Hammer:2015pcx,Babichev:2024eoh} (see also Ref.~\cite{Domenech:2023ryc} for a generalization). Conformal symmetry also relates mimetic gravity with general relativistic generalizations of MOND \cite{Domenech:2025qny}. For recent reviews of mimetic gravity see Refs.~\cite{Sebastiani:2016ras,Malaeb:2026ecv}.

The Paneitz operator, a dimension-4 conformally invariant operator, applied to dimension-zero scalar fields has attracted attention for possible cancellations of the vacuum energy and generation of a scale-invariant primordial spectrum \cite{Boyle:2021jaz,Turok:2023amx}. Recently, however, Ref.~\cite{Cline:2026zcv} has shown that the theory suffers from ghosts and is, therefore, not viable. Here we show that, owing to its conformal invariant nature, the Paneitz operator in fact corresponds to a generalized class of mimetic gravity, so-called extended mimetic gravity \cite{Takahashi:2017pje}.

Since its inception \cite{Chamseddine:2013kea,Chamseddine:2014vna,Chaichian:2014qba,Mirzagholi:2014ifa}, mimetic gravity has been expanded to the most general degenerate higher order scalar tensor (DHOST) theories \cite{Langlois:2018jdg} (see \cite{Langlois:2015cwa,BenAchour:2016cay,BenAchour:2016fzp,Crisostomi:2016czh,deRham:2016wji} for original works on DHOST), including studies of cosmological perturbations. See also Refs.~\cite{Ganz:2018mqi,Ganz:2019vre} for Hamiltonian analyses and Ref.~\cite{Babichev:2024eoh} for further generalizations. It is well-known that general mimetic gravity suffers from either ghosts or gradient instabilities \cite{Langlois:2018jdg,Firouzjahi:2017txv,Hirano:2017zox,Takahashi:2017pje,Zheng:2017qfs,Gorji:2017cai}.  These results, therefore, apply to the Paneitz operator as well. Nevertheless, a potential solution involves adding higher curvature corrections \cite{Gorji:2017cai,Zheng:2017qfs,Hirano:2017zox}.

In this work, we revisit the Paneitz operator in conformal-invariant scalar-tensor theories and show that it is a mimetic DHOST theory, complementing Ref.~\cite{Cline:2026zcv}. Since the Paneitz operator alone has instabilities in the tensor sector \cite{Cline:2026zcv}, we add the Einstein-Hilbert term in the Lagrange multiplier formulation of mimetic gravity \cite{Golovnev:2013jxa,Arroja:2015wpa}. This ensures that the ``Einstein-Hilbert term'' is compatible with the conformal symmetry of the original action. We find that even in that case, the Paneitz operator is strongly constrained by modifications to the propagation speed of gravitational waves (GWs) \cite{LIGOScientific:2017zic}, assuming higher curvature corrections stabilize the scalar sector.

This paper is organized as follows. In \S.~\ref{sec:1}, we show that the Paneitz action corresponds to mimetic DHOST and discuss the coupling to matter. In \S.~\ref{sec:2}, we study its cosmology, from background to linear perturbations, also including the standard Einstein-Hilbert term. We derive constraints on the Paneitz operator coefficient from modifications to GW propagation speed. Lastly, we end our work in \S.~\ref{sec:3} with further discussions.

\section{Paneitz action and Mimetic gravity \label{sec:1}}

The Paneitz operator \cite{Paneitz:2008afy} is a conformal invariant fourth-order differential operator, originally motivated by conformal supergravity \cite{Kaku:1978nz,Fradkin:1981jc,Fradkin:1982xc}, that we denote as $\Delta_4$ and is explicitly given by
\begin{align}\label{eq:Paneitz-operator}
    \Delta_4=\Box^2  +2\nabla_{\mu}\left[\left(R{^{\mu\nu}}-\frac{1}{3}g^{\mu\nu}R\right)\nabla_{\nu}\right]\,,
\end{align} 
where $R{^{\mu\nu}}$ and $R$ are the Ricci tensor and Ricci scalar respectively. Such an operator is only rescaled after a conformal transformations of the metric. Namely, under a metric transformation given by
\begin{equation}\label{eq:conformal-transf}
    g_{\mu\nu}\rightarrow\Omega^2(x^\mu)g_{\mu\nu},
\end{equation}
where $\Omega$ is a smooth, strictly positive function, the Paneitz operator correspondingly scales as $\Delta_4\to \Omega^{-4} \Delta_4$. Thus, one may build a conformal invariant action for a dimension-zero scalar field, say $\phi$, given by
\begin{align}\label{eq:Paneitz-action}
    S_{\Delta_4}=\int \dd^4x\,\sqrt{-g}\, \phi \Delta_4 \phi\,.
\end{align}
Eq.~\eqref{eq:Paneitz-action} is trivially conformal invariant since $\sqrt{-g}\to \Omega^4 \sqrt{-g}$ after the conformal transformation, cancelling the rescaling of the Paneitz operator.
Throughout this work, we refer to the action \eqref{eq:Paneitz-action} as the \textit{Paneitz action}. 

\subsection{Connection to mimetic DHOST}

The action Eq.~\eqref{eq:Paneitz-action} seemingly has higher derivatives stemming from the Paneitz operator \eqref{eq:Paneitz-operator}, which could lead to the so-called Ostrogradsky ghosts \cite{Woodard:2015zca}. The conformal symmetry, however, introduces a first-class constraint and prevents such a ghost like in standard mimetic gravity \cite{Ganz:2018mqi,Ganz:2019vre}, see also Ref.~\cite{BenAchour:2016cay}.

To see this and the connection to mimetic gravity, let us rewrite Eq.~\eqref{eq:Paneitz-action}, after integration by parts, as
\begin{align}\label{eq:Paneitz-Lagrangian}
     S_{\Delta_4}=\int\dd^4x\sqrt{-g}\left\{\frac{2X}{3}R -\left(\Box\phi\right)^2+2\nabla_{\mu}\nabla_{\nu}\phi\nabla^{\mu}\nabla^{\nu}\phi\right\}\,,
\end{align}
where we introduced $X=\nabla_{\mu}\phi\nabla^{\mu}\phi$. In this form, we identify that the Paneitz action \eqref{eq:Paneitz-Lagrangian} belongs to the class of quadratic DHOST models \cite{Langlois:2015cwa}. More concretely, in the notation of  \cite[Section~IV]{Langlois:2018dxi} this corresponds to $F=2X/3$, $A_1=2$, $A_2=-1$ and $A_3=A_4=A_5=0$.

The conformal invariance of the original action \eqref{eq:Paneitz-action} can be made more apparent by expressing Eq.~\eqref{eq:Paneitz-Lagrangian} in terms of conformally invariant tensors as in, e.g., Ref.~\cite{Domenech:2023ryc,Babichev:2024eoh}. Thus, we introduce a conformal invariant metric,
\begin{align}\label{eq:InvariantMetric}
    {\cal G}_{\mu\nu} = X g_{\mu\nu}\,,
\end{align}
and conformally invariant second derivative and curvature tensors, respectively given by
\begin{align}\label{eq:InvariantExpressions}
{\cal C}_{\mu\nu} =  \nabla^{\cal G}_\mu \nabla^{\cal G}_\nu \phi\quad,\quad{\cal R}_{\mu\nu\alpha\beta} = & R[{\cal G}]_{\mu\nu\alpha\beta}\,,
\end{align}
where $\nabla^{\cal G}_\mu $ and $R[{\cal G}]_{\mu\nu\alpha\beta}$ mean the covariant derivative and the Riemann tensor associated with the invariant metric \eqref{eq:InvariantMetric}. One can in principle build any conformal invariant models out of these tensors and its derivative, and the resulting theory belongs to mimetic gravity \cite{Babichev:2024eoh}.

In terms of  the conformal invariant objects \eqref{eq:InvariantMetric} and \eqref{eq:InvariantExpressions}, the Paneitz action \eqref{eq:Paneitz-action} now reads
\begin{align}\label{eq:InvariantPaneitz}
   S_{\Delta_4} = \int   \dd^4x&\sqrt{- {\cal G}}\bigg\{ \frac{2}{3} {\cal G}^{\alpha\mu}{\cal G}^{\beta\nu}{\cal R}_{\alpha\beta\mu\nu} \nonumber\\ 
   &\,- ({\cal G}^{\alpha\beta} C_{\alpha\beta})^2+2 {\cal G}^{\alpha\mu}{\cal G}^{\beta\nu}{\cal C}_{\alpha\beta}{\cal C}_{\mu\nu}\bigg\}\,,
\end{align}
which is manifestly conformally invariant.
At this point, we have arrived at a formulation which is exactly like the original mimetic gravity  mimetic gravity by using conformally invariant auxiliary metric \cite{Chamseddine:2013kea,Jirousek:2022rym}.  However, we can make the correspondence with mimetic gravity more explicit. 

The action Eq.~\eqref{eq:InvariantPaneitz} contains a conformal gauge degree of freedom which can be fixed. For instance, one can promote the field $X$ to a new field, say $\chi$, via a Lagrange multiplier. Assuming for the moment that $X\neq 0$, one can choose a conformal frame transformation such that $\chi=\pm 1$ depending on whether the scalar field is timelike or spacelike. By doing so, we arrive at  \cite{Golovnev:2013jxa,Arroja:2015wpa}
\begin{align}\label{eq:paneitzgaugefixed}
    S_{\Delta_4} = \int \dd^4x\sqrt{-g}\,  &\bigg\{ \mp \frac{2}{3} R +2 \nabla_\mu \nabla_\nu \phi \nabla^\mu \nabla^\nu \phi\nonumber\\ 
    &\quad\quad\quad - (\Box \phi)^2 + \lambda (X \pm 1 ) \bigg\}\,.
\end{align}
Variation of \eqref{eq:paneitzgaugefixed} with respect to the Lagrange multiplier $\lambda$ yields the \textit{mimetic constraint}, namely
\begin{align}\label{eq:mimetic_constraint}
    X=\nabla_\mu \phi \nabla^\mu \phi =\pm 1 \,.
\end{align}
Note that we can use the constraint \eqref{eq:mimetic_constraint} directly into the action \eqref{eq:paneitzgaugefixed}, except for the terms inside the Lagrange multiplier, because the mimetic constraint completely fixes the conformal gauge \cite{Motohashi:2016prk}.

Mimetic gravity actions of this form have been studied in detail in \cite[Section~4]{Langlois:2018jdg} (see also Refs.~\cite{Firouzjahi:2017txv,Hirano:2017zox,Takahashi:2017pje,Zheng:2017qfs,Gorji:2017cai}), and found that these models generically suffer from ghost or gradient instabilities for linear perturbations around the flat FLRW background. We will explicitly check this for the Paneitz operator later in \S~\ref{sec:2}. Note that in the original mimetic dark matter formulation \cite{Chamseddine:2013kea}, the gradient instability corresponds to the standard Jeans instability of a pressure-less fluid. The analogy no longer holds for the general mimetic DHOST case. Nevertheless, such instabilities can be avoided by adding higher curvature terms to the original action \cite{Gorji:2017cai,Zheng:2017qfs,Hirano:2017zox}.

In passing, it is interesting to note that the conformally invariant operators \eqref{eq:InvariantExpressions} and the conformal transformation \eqref{eq:InvariantMetric} are only well defined if $X\neq 0$, that is, $X$ is either spacelike or timelike. However, the Paneitz action Eq. \eqref{eq:Paneitz-Lagrangian} is a priori well-defined for the null-like case, that is, $X=0$. Indeed, in flat spacetime, one of the solutions is given by $\phi = \phi(x^\mu k_\mu)$ with $k_\mu k^\mu =0$. 
This solution is valid for any conformal flat metric $g_{\mu\nu} = \Omega(x^\alpha) \eta_{\mu\nu}$ and, therefore, for flat Friedmann–Lemaître–Robertson–Walker (FLRW) metrics in conformal time. Note that as soon as additional matter is included, this solution breaks down (see \S.~\ref{sec:mattercoupling}).

A null-like gradient may resolve common issues with black hole solutions in the original mimetic gravity models \cite{Myrzakulov:2015kda,Gorji:2020ten}, allowing a smooth transition from spacelike to timelike field gradients.
We checked that for a shift-symmetric conformally invariant theory up to second derivatives of the scalar field and the metric, the Paneitz action \eqref{eq:Paneitz-action} and the K-essence model with $L = \sqrt{-g} X^2$, which is also conformal invariant, are the only ones consistent with a smooth transition from spacelike to timelike gradients of the scalar field. Note that, when $X \neq 0$, the K-essence term is equivalent to a cosmological constant after gauge fixing the conformal symmetry. We leave a detailed study of the null-like case for future work.

\subsection{Matter coupling \label{sec:mattercoupling}}

Let us briefly discuss how to couple matter to the Paneitz action and the mimetic field $\phi$. Note that if we naively add the conformally invariant action $S_{\Delta_4}$ 
\eqref{eq:Paneitz-action} to the standard matter sector, say $S_M$, the resulting theory breaks the conformal invariance and, thereby, revives the Ostrogradsky ghost instability. To avoid this issue, one can either couple matter directly to the conformally invariant metric ${\cal G}_{\mu\nu}$, that is
\begin{align}
    S = - \alpha S_{\Delta_4} + S_M({\cal G}_{\mu\nu},\Psi)\,,
\end{align}
where $\Psi$ refers to matter fields,
or fix the conformal symmetry beforehand by introducing, for instance, the mimetic constraint \eqref{eq:mimetic_constraint}, namely
\begin{align}\label{eq:gaugefixedmatter}
    S = &- \alpha S_{\Delta_4} + S_M(g_{\mu\nu},\Psi) +\int \sqrt{-g} \lambda ( X \pm 1)\mathrm{d}^4x\,,
\end{align}
where $\alpha$ is a constant coefficient and the $-$ sign in front is for later convenience. 
Both these approaches are equivalent for $X \neq 0$ up to conformal gauge fixing.  Note that for standard model particles, the limit $X \rightarrow 0$ shuts off the matter coupling for fermions and the Higgs field (see, e.g., Ref.~\cite{Domenech:2025gao} for disformal couplings to fermions) and, therefore, this limit is only relevant for the conformally invariant matter sector.

\section{Cosmology of Paneitz action  \label{sec:2}}

In this section, we look at the Paneitz action \eqref{eq:Paneitz-action} in an FRLW universe. We first examine the background dynamics without conformal gauge fixing to show how conformal symmetry leads to a dust-like component in the Friedmann equations. To consider a more realistic model with a standard General Relativity limit, we then add the conformal invariant Ricci scalar, that is, the standard Einstein-Hilbert term in the conformal gauge fixed formulation, and study the FLRW background and linear cosmological perturbations.

\subsection{Paneitz action with no gauge fixing}

Consider the Paneitz action \eqref{eq:Paneitz-action} in a flat FLRW background given by 
\begin{align}\label{eq:frlwbg}
    \dd s^2 = - N^2 \dd t^2 + a^2 \delta_{ij} \dd x^i \dd x^j\,,
\end{align}
and a homogeneous scalar field $\phi=\phi_0(t)$. Inserting these ansatze in Eq.~\eqref{eq:Paneitz-action}, we arrive at
\begin{align}\label{eq:paneitzflrwbg}
    S_{\Delta_4} = - \alpha \int \dd^4x\, \frac{a^3}{N^3}   \left( \frac{N \dot a - a \dot N}{a N} \dot \phi_0 + \ddot \phi_0  \right)^2\,.
\end{align}
To make the symmetries more apparent, we rewrite Eq.~\eqref{eq:paneitzflrwbg} equivalently as
\begin{align}\label{eq:paneitzflrwbg2}
    S_{\Delta_4} = - \alpha \int \dd^3x \,(\dot\phi\,\dd t)\, \left(\frac{a\dot\phi}{N}\right)^3   \left(\frac{d}{\dot\phi dt}\left[\frac{a\dot\phi}{N}\right]\right)^2\,.
\end{align}
From Eq.~\eqref{eq:paneitzflrwbg2} the conformal symmetry is clear as it corresponds to $a\to \Omega a$ and $N\to \Omega N$. It is also clear that the only dynamical variable is the combination $a\dot\phi/N$. Note that, outside of such combination, $\dot\phi$ plays the role of the standard lapse in the action \eqref{eq:paneitzflrwbg2}, as it appears always in the combination $\dot\phi dt$, and can, therefore, be set to $\dot\phi=1$, in Planck units, by time reparametrization invariance. The conformal symmetry may be used to set $N=1$, thus fixing all gauge degrees of freedom. This is equivalent to imposing the mimetic constraint \eqref{eq:mimetic_constraint} from the start and setting $N=1$, as is standard practice.

Let us for now proceed without gauge fixing. To include a consistent matter field, we consider a radiation fluid, which is conformal invariant. Adding the radiation fluid, we obtain that the equations of motion in conformal time, that is we set $N = a$, are given bye
\begin{align}\label{eq:1stfried}
   \alpha \frac{\phi_0^{\prime\prime 2} - 2 \phi_0^\prime \phi_0^{\prime\prime\prime} }{2 a^2} = \rho a^2\,, \\
    \alpha \frac{\phi_0^{\prime\prime 2} - 2 \phi_0^\prime \phi_0^{\prime\prime\prime} }{6 a^2} = p a^2\,,\label{eq:2ndfried}
\end{align}
where $\rho$ and $p$ are the energy density and pressure of radiation.   
On the other hand, the equations of motion for the scalar field lead to a conserved Noether current, that is $\nabla_\mu J^\mu=0$ where  $J^\mu$ reads
\begin{align}
    J^\mu = - 2 \alpha \frac{\phi_0^{\prime\prime\prime}}{a^4} \delta^\mu_0
\end{align}
Integrating the conserved current leads to a dark matter-like contribution from the integration constant. Namely, after integrating the scalar field equation, we get from Eq.~\eqref{eq:1stfried} that
\begin{align}
   \alpha \frac{\phi_0^{\prime\prime}}{2 a^2} = \rho a^2 + \frac{\phi_0^\prime}{a}\rho_{\rm DM} a^2 \,,
\end{align}
where we defined $- 2 \alpha \phi_0^{\prime\prime\prime} = {\mathrm{constant}} = \rho_{\rm DM} a^3$. Fixing the conformal symmetry via $\phi_0^\prime=a$ and rescaling $\alpha$, we recover the standard first Friedmann equation.

\subsection{Einstein-Hilbert plus Paneitz actions}

We now include study a model with the standard general relativity limit. Note that we cannot add the standard Einstein-Hilbert action directly into the Paneitz action \eqref{eq:Paneitz-action}, in the conformal invariant formulation, because it would spoil the conformal invariance and revive the Ostrogradsky ghost. Instead, we must add the conformal invariant Ricci scalar using Eqs.~\eqref{eq:InvariantMetric} and \eqref{eq:InvariantExpressions}, i.e., the original mimetic matter model \cite{Chamseddine_2013}, namely
\begin{align}
    S =& \int \dd^4x\, \sqrt{-g} \Big[  - \alpha_{\rm GR} X^2  {\cal R}[{\cal G}] - \alpha_{\rm Pan} \phi \Delta_4 \phi   \Big] \nonumber \\
    =& \int \dd^4x\, \sqrt{-g} \Big[ - \left( \alpha_{\rm GR} + \frac{2}{3} \alpha_{\rm Pan} \right) X R + \alpha_{\rm Pan} (\Box \phi)^2 \nonumber \\
    & - 2 \alpha_{\rm Pan}  \nabla_\mu \nabla_\nu \phi \nabla^\mu \nabla^\nu \phi - \frac{3 \alpha_{\rm GR} }{2 X} \nabla_\mu X \nabla^\mu X\Big]\,,
\end{align}
where in the last step we explicitly wrote the action in terms of derivatives of $\phi$ and the Ricci scalar of the metric $g_{\mu\nu}$, after integration by parts. We also introduced $\alpha_{\rm GR}$ and $\alpha_{\rm Pan}$ to denote the coefficients in front of the conformal Ricci scalar, which later lead to the standard Einstein-Hilbert term plus the mimetic constraint, and in front of the Paneitz operator, respectively. 

Assuming $X  \neq 0$, we can fix the conformal gauge by fixing $X = - \Lambda^2$ via a Lagrange multiplier, where we introduced an energy scale $\Lambda$ to keep track of the units. Note that we consider $X<0$ because we are interested in the cosmological setup. This leads us to the equivalent action given by
\begin{align}\label{eq:fullactionfixed}
    S = & \alpha_{\rm GR} \Lambda^2 \int \dd^4x\, \Big[ \left(  \frac{2 \alpha_{\rm Pan}}{3 \alpha_{\rm GR}} + 1 \right) R + \frac{\alpha_{\rm Pan}}{\alpha_{\rm GR} \Lambda^2} (\Box \phi)^2 \nonumber \\
    & - \frac{2 \alpha_{\rm Pan} }{\alpha_{\rm GR} \Lambda^2} \nabla_\mu \nabla_\nu \phi \nabla^\mu \nabla^\nu \phi + \lambda (X + \Lambda^2) \Big]\,.
\end{align}
The equations of motion that follow from the variation of Eq.~\eqref{eq:fullactionfixed} in the flat FLRW background \eqref{eq:frlwbg} are given by
\begin{align}\label{eq:1stfriedfull}
   3\alpha_{\rm GR} \Lambda^2 \left( 1 + \frac{ \alpha_{\rm Pan}}{6 \alpha_{\rm GR} }  \right) H^2 = \rho +  \frac{C_1 }{ a^3  }, \\
   \left(1+ \frac{\alpha_{\rm Pan}}{6 \alpha_{\rm GR} } \right)  \alpha_{\rm GR} \Lambda^2 \left( 3 H^2 + 2 \dot H \right) = p \,,\label{eq:2ndfriedfull}
\end{align}
where $C_1$ is an integration constant, we used $\dot \phi =\Lambda$ due to the gauge-fixing constraint, and we used that from the scalar field one obtains
\begin{align}
   \lambda_0 = - \frac{C_1}{a^3 \Lambda^4 \alpha_{\rm GR} } - \frac{3 \alpha_{\rm Pan}}{\alpha_{\rm GR} \Lambda^2 } \left( 2 H^2 + \dot H \right)\,.
\end{align}
Note that in Eq.~\eqref{eq:1stfriedfull} we considered an additional perfect fluid with energy density $\rho$ and pressure $p$. We can do that without worrying about the matter coupling since we are already working with the conformal gauge fixed formulation in Eq.~\eqref{eq:fullactionfixed}, see also Eq.~\eqref{eq:gaugefixedmatter}.

From Eqs.~\eqref{eq:1stfriedfull} and \eqref{eq:2ndfriedfull}, we see that we recover the standard Friedmann equations in a flat FLRW background with a perfect fluid and a dust fluid (from the integration constant $C_1$,) except for the common prefactor $\alpha_{\rm GR} \Lambda^2 \left( 1 + { \alpha_{\rm Pan}}/({6 \alpha_{\rm GR} })  \right)$, which could be fixed to be the standard $M_{\rm pl}^2=1/\sqrt{8\pi G}$ in natural units. We keep this prefactor arbitrary for the moment, as our conclusions do not depend on the value of $\Lambda$.

\subsubsection{Gravitational waves}

Since the background equations of motion are basically the standard Friedmann equations with an additional dust fluid, we look at the behavior of cosmological perturbations. For simplicity, we study first the tensor fluctuations and later turn to scalar ones. To study tensor fluctuations we perturb the flat FLRW metric as
\begin{align}
   \mathrm{d}s^2 = a(\tau)^2 ( - \mathrm{d}t^2 + (\delta_{ij} + h_{ij} \mathrm{d}x^i \mathrm{d}x^j) )\,,
\end{align}
where $h_{ij}$ are the traceless and transverse tensor modes associated with GWs. After perturbative expansion of the action \eqref{eq:fullactionfixed}, and after some calculations, we find that second order action for the tensor modes reads
\begin{align}\label{eq:Sh}
    S_h = \alpha_{\rm GR}\Lambda^2 \int \dd^4x &\frac{a^3}{4} \bigg\{ \left( 1 - \frac{4\alpha_{\rm Pan}}{3\alpha_{\rm GR}}  \right) \dot h_{ij}^2 \nonumber  \\&
    \quad- \left( 1 + \frac{2\alpha_{\rm Pan}}{3\alpha_{\rm GR}} \right) (\partial_k h_{ij})^2 \bigg\}\,.
\end{align}
From Eq.~\eqref{eq:Sh}, we see that the propagation speed of the GWs is given by
\begin{align}\label{eq:cg}
    c_g^2= \frac{3\alpha_{\rm GR} + 2 \alpha_{\rm Pan}}{3\alpha_{\rm GR} - 4 \alpha_{\rm Pan} }\,.
\end{align}
We have checked that our result agrees with Ref.~\cite{Langlois:2018jdg}.

It is interesting to note that tensor fluctuations are unstable when $\alpha_{\rm GR}=0$ as the time derivative and gradient terms have a same relative sign, leading to either ghost or gradient instabilities for $\alpha_{\rm Pan}<0$ and $\alpha_{\rm Pan}>0$, respectively. This was already pointed out by Ref.~\cite{Cline:2026zcv}. However, by including the conformal Ricci scalar term, tensor modes can be stable. In particular, from Eq.~\eqref{eq:cg} we see that $c_g=1$ when $\alpha_{\rm Pan}\to 0$. In general, the tensor modes in Eq.~\eqref{eq:Sh} are stable if
\begin{align}\label{eq:stabilityh}
\alpha_{\rm GR}>0 \quad {\rm and}\quad -\frac{3}{2}\alpha_{\rm GR}<\alpha_{\rm Pan}<\frac{3}{4}\alpha_{\rm GR}\,.
\end{align}

For $\alpha_{\rm Pan}\neq 0$, we can use Eq.~\eqref{eq:cg} to place severe constraints on the coefficient of the Paneitz operator, $\alpha_{\rm Pan}$.  In fact, from the simulatenous observation of a GW event and the electromagnetic counterpart of a neutron star merger \cite{LIGOScientific:2017zic,Ezquiaga:2017ekz}, the propagation speed of gravitational waves $c_g$ in that frequency range has been constrained to be
\begin{equation}\label{eq.observation}
     \left|{c_g}-1\right|\leq 5\times 10^{-16}\,,
\end{equation}
where we set the speed of light $c=1$.
Using Eq.~\eqref{eq:cg}, we find that
\begin{align}\label{eq:constrain}
|\alpha_{\rm Pan}|/\alpha_{\rm GR}<5\times 10^{-16}\,,
\end{align}
and thus the Paneitz operator must have a negligible effect at least in the recent Universe.

\subsubsection{Scalar sector}

For completeness, we also study the scalar fluctuations. Working in the uniform-$\phi$ slicing, also called the unitary gauge, we perturb the FLRW metric as
\begin{align}
    \dd s^2 = - e^{2 \Phi} \dd t^2 + a^2 e^{2\zeta} \delta_{ij} ( \partial^i \beta  + \dd x^i  ) ( \partial^j \beta  + \dd x^i  ) \,.
\end{align}
Inserting this ansatz into the action \eqref{eq:fullactionfixed}, Taylor expanding, solving the constraints, and after some calculations, we find that the second order action for the curvature perturbation $\zeta$ is given by
\begin{align}\label{eq:actionforzeta}
    S =2 \alpha_{\rm GR} \Lambda^2\int \dd^4x\, a^3 &\bigg\{ \left(\frac{1}{3}+\frac{\alpha_{\rm GR}}{\alpha_{\rm Pan}}-\frac{20}{9}\frac{\alpha_{\rm Pan}}{\alpha_{\rm GR}}\right) \dot \zeta^2  \nonumber \\&
    + \left(1+\frac{2 \alpha_{\rm Pan} }{3 \alpha_{\rm GR} }\right)   (\partial_k \zeta)^2  \bigg\}\,,
\end{align}
where we have assumed that $\alpha_{\rm Pan} \neq 0$. This result is in agreement with Ref.~\cite{Langlois:2018jdg}. Note that, in Eq.~\eqref{eq:actionforzeta}, one cannot take the limit $\alpha_{\rm Pan} \to0$ after solving the constraints, as it would be inconsistent with the simplifications. However, if we set  $\alpha_{\rm Pan} =0$ from the start, we recover the standard mimetic dark matter case \cite{Chamseddine:2014vna}. 

Stability of the tensor modes requires the condition \eqref{eq:stabilityh}. However, if $\alpha_{\rm GR}>0$, stability of $\zeta$ \eqref{eq:actionforzeta} requires $\alpha_{\rm Pan}<-3\alpha_{\rm GR}/2$, inconsistent with Eq.~\eqref{eq:stabilityh}. Thus, one either has a ghost or gradient instability in $\zeta$ or $h_{ij}$. This is a general result of mimetic gravity as discussed in details in Ref.~\cite{Takahashi:2017pje,Langlois:2018jdg,Firouzjahi:2017txv,Hirano:2017zox,Gorji:2017cai,Zheng:2017qfs}. Such instability can be avoided via a direct coupling between the higher-derivative operators and Ricci curvature \cite{Hirano:2017zox,Gorji:2017cai,Zheng:2017qfs}. Unfortunately, as we have shown, even with higher-derivative terms, the presence of the Paneitz operator is severely constrained by GW observations \eqref{eq:constrain}. Thus, the model of \cite{Boyle:2021jaz,Turok:2023amx} is at odds with observations in its simplest, original form.

\section{Conclusions and discussion \label{sec:3}}

We have shown that the Paneitz action \eqref{eq:Paneitz-action}, which is a conformal invariant action involving a dimension-zero scalar field and a fourth-order differential operator, the Paneitz operator, is equivalent to extended mimetic gravity theories and falls within mimetic DHOST gravity \cite{Langlois:2018jdg}. As such, linear cosmological perturbations have a ghost or gradient instability in either the tensor or scalar sector \cite{Takahashi:2017pje,Langlois:2018jdg,Firouzjahi:2017txv,Hirano:2017zox,Gorji:2017cai,Zheng:2017qfs}. Although unappealing, this is not a problem per se, as it can be cured by adding higher-derivative corrections \cite{Hirano:2017zox,Gorji:2017cai,Zheng:2017qfs}. One of the main obstacles, however, is that the propagation speed of GWs is modified, which is severely constraint by current observations \cite{LIGOScientific:2017zic}. Thus, the Paneitz operator must have a very small coefficient today, see Eq.~\eqref{eq:constrain}.

Such a model attracted recent attention for the possible cancellation of vacuum energy and explanation of primordial fluctuations \cite{Boyle:2021jaz,Turok:2023amx}. However, it has been deemed not viable by Ref.~\cite{Cline:2026zcv}. Our work complements that of Ref.~\cite{Cline:2026zcv} by embedding the Paneitz operator applied to a dimension-zero scalar field within mimetic gravity. Our insight shows that the Paneitz action \eqref{eq:Paneitz-action} can thus be consistently added the conformal invariant formulation of mimetic gravity, with a consistent matter coupling (see \S.~\ref{sec:mattercoupling}). Otherwise, the theory contains an Ostrogradsky ghost.

Our work also provides potential, consistent extensions of the proposal by Refs.~\cite{Boyle:2021jaz,Turok:2023amx} within more general mimetic DHOST models \cite{Langlois:2018jdg} and with higher-derivative corrections \cite{Hirano:2017zox,Gorji:2017cai,Zheng:2017qfs} to render the theory stable and with $c_g=1$ today. It would be interesting to revisit vacuum cancellations and primordial fluctuations within such a consistent model. We note, however, that adding multiple dimension-zero scalar fields with the Paneitz operator likely results in the presence of several Ostrogradsky ghosts, as the conformal symmetry only eliminates one of the ghost degrees of freedom. We leave a detailed study of such a possibility for future work.

\section*{Acknowledgements}
We thank Anamaria Hell for useful correspondence. This research is supported by the DFG under the Emmy-Noether program, project number 496592360, and by the JSPS KAKENHI grant No. JP24K00624.
\newpage
\bibliography{thebibliography.bib}

%apsrev4-2.bst 2019-01-14 (MD) hand-edited version of apsrev4-1.bst
%Control: key (0)
%Control: author (8) initials jnrlst
%Control: editor formatted (1) identically to author
%Control: production of article title (0) allowed
%Control: page (0) single
%Control: year (1) truncated
%Control: production of eprint (0) enabled
\begin{thebibliography}{67}%
\makeatletter
\providecommand \@ifxundefined [1]{%
 \@ifx{#1\undefined}
}%
\providecommand \@ifnum [1]{%
 \ifnum #1\expandafter \@firstoftwo
 \else \expandafter \@secondoftwo
 \fi
}%
\providecommand \@ifx [1]{%
 \ifx #1\expandafter \@firstoftwo
 \else \expandafter \@secondoftwo
 \fi
}%
\providecommand \natexlab [1]{#1}%
\providecommand \enquote  [1]{``#1''}%
\providecommand \bibnamefont  [1]{#1}%
\providecommand \bibfnamefont [1]{#1}%
\providecommand \citenamefont [1]{#1}%
\providecommand \href@noop [0]{\@secondoftwo}%
\providecommand \href [0]{\begingroup \@sanitize@url \@href}%
\providecommand \@href[1]{\@@startlink{#1}\@@href}%
\providecommand \@@href[1]{\endgroup#1\@@endlink}%
\providecommand \@sanitize@url [0]{\catcode `\\12\catcode `\$12\catcode
  `\&12\catcode `\#12\catcode `\^12\catcode `\_12\catcode `\%12\relax}%
\providecommand \@@startlink[1]{}%
\providecommand \@@endlink[0]{}%
\providecommand \url  [0]{\begingroup\@sanitize@url \@url }%
\providecommand \@url [1]{\endgroup\@href {#1}{\urlprefix }}%
\providecommand \urlprefix  [0]{URL }%
\providecommand \Eprint [0]{\href }%
\providecommand \doibase [0]{https://doi.org/}%
\providecommand \selectlanguage [0]{\@gobble}%
\providecommand \bibinfo  [0]{\@secondoftwo}%
\providecommand \bibfield  [0]{\@secondoftwo}%
\providecommand \translation [1]{[#1]}%
\providecommand \BibitemOpen [0]{}%
\providecommand \bibitemStop [0]{}%
\providecommand \bibitemNoStop [0]{.\EOS\space}%
\providecommand \EOS [0]{\spacefactor3000\relax}%
\providecommand \BibitemShut  [1]{\csname bibitem#1\endcsname}%
\let\auto@bib@innerbib\@empty
%</preamble>
\bibitem [{\citenamefont {Wetterich}(2019)}]{Wetterich:2019qzx}%
  \BibitemOpen
  \bibfield  {author} {\bibinfo {author} {\bibfnamefont {C.}~\bibnamefont
  {Wetterich}},\ }\bibfield  {title} {\bibinfo {title} {{Quantum scale
  symmetry}},\ }\href@noop {} {\  (\bibinfo {year} {2019})},\ \Eprint
  {https://arxiv.org/abs/1901.04741} {arXiv:1901.04741 [hep-th]} \BibitemShut
  {NoStop}%
\bibitem [{\citenamefont {Wetterich}(2021)}]{Wetterich:2020cxq}%
  \BibitemOpen
  \bibfield  {author} {\bibinfo {author} {\bibfnamefont {C.}~\bibnamefont
  {Wetterich}},\ }\bibfield  {title} {\bibinfo {title} {{Fundamental scale
  invariance}},\ }\href {https://doi.org/10.1016/j.nuclphysb.2021.115326}
  {\bibfield  {journal} {\bibinfo  {journal} {Nucl. Phys. B}\ }\textbf
  {\bibinfo {volume} {964}},\ \bibinfo {pages} {115326} (\bibinfo {year}
  {2021})},\ \Eprint {https://arxiv.org/abs/2007.08805} {arXiv:2007.08805
  [hep-th]} \BibitemShut {NoStop}%
\bibitem [{\citenamefont {Wetterich}(1984)}]{Wetterich:1983bi}%
  \BibitemOpen
  \bibfield  {author} {\bibinfo {author} {\bibfnamefont {C.}~\bibnamefont
  {Wetterich}},\ }\bibfield  {title} {\bibinfo {title} {{Fine Tuning Problem
  and the Renormalization Group}},\ }\href
  {https://doi.org/10.1016/0370-2693(84)90923-7} {\bibfield  {journal}
  {\bibinfo  {journal} {Phys. Lett. B}\ }\textbf {\bibinfo {volume} {140}},\
  \bibinfo {pages} {215} (\bibinfo {year} {1984})}\BibitemShut {NoStop}%
\bibitem [{\citenamefont {Hempfling}(1996)}]{Hempfling:1996ht}%
  \BibitemOpen
  \bibfield  {author} {\bibinfo {author} {\bibfnamefont {R.}~\bibnamefont
  {Hempfling}},\ }\bibfield  {title} {\bibinfo {title} {{The Next-to-minimal
  Coleman-Weinberg model}},\ }\href
  {https://doi.org/10.1016/0370-2693(96)00446-7} {\bibfield  {journal}
  {\bibinfo  {journal} {Phys. Lett. B}\ }\textbf {\bibinfo {volume} {379}},\
  \bibinfo {pages} {153} (\bibinfo {year} {1996})},\ \Eprint
  {https://arxiv.org/abs/hep-ph/9604278} {arXiv:hep-ph/9604278} \BibitemShut
  {NoStop}%
\bibitem [{\citenamefont {Meissner}\ and\ \citenamefont
  {Nicolai}(2007)}]{Meissner:2006zh}%
  \BibitemOpen
  \bibfield  {author} {\bibinfo {author} {\bibfnamefont {K.~A.}\ \bibnamefont
  {Meissner}}\ and\ \bibinfo {author} {\bibfnamefont {H.}~\bibnamefont
  {Nicolai}},\ }\bibfield  {title} {\bibinfo {title} {{Conformal Symmetry and
  the Standard Model}},\ }\href
  {https://doi.org/10.1016/j.physletb.2007.03.023} {\bibfield  {journal}
  {\bibinfo  {journal} {Phys. Lett. B}\ }\textbf {\bibinfo {volume} {648}},\
  \bibinfo {pages} {312} (\bibinfo {year} {2007})},\ \Eprint
  {https://arxiv.org/abs/hep-th/0612165} {arXiv:hep-th/0612165} \BibitemShut
  {NoStop}%
\bibitem [{\citenamefont {Foot}\ \emph {et~al.}(2007)\citenamefont {Foot},
  \citenamefont {Kobakhidze},\ and\ \citenamefont {Volkas}}]{Foot:2007as}%
  \BibitemOpen
  \bibfield  {author} {\bibinfo {author} {\bibfnamefont {R.}~\bibnamefont
  {Foot}}, \bibinfo {author} {\bibfnamefont {A.}~\bibnamefont {Kobakhidze}},\
  and\ \bibinfo {author} {\bibfnamefont {R.~R.}\ \bibnamefont {Volkas}},\
  }\bibfield  {title} {\bibinfo {title} {{Electroweak Higgs as a
  pseudo-Goldstone boson of broken scale invariance}},\ }\href
  {https://doi.org/10.1016/j.physletb.2007.06.084} {\bibfield  {journal}
  {\bibinfo  {journal} {Phys. Lett. B}\ }\textbf {\bibinfo {volume} {655}},\
  \bibinfo {pages} {156} (\bibinfo {year} {2007})},\ \Eprint
  {https://arxiv.org/abs/0704.1165} {arXiv:0704.1165 [hep-ph]} \BibitemShut
  {NoStop}%
\bibitem [{\citenamefont {Aoki}\ and\ \citenamefont {Iso}(2012)}]{Aoki:2012xs}%
  \BibitemOpen
  \bibfield  {author} {\bibinfo {author} {\bibfnamefont {H.}~\bibnamefont
  {Aoki}}\ and\ \bibinfo {author} {\bibfnamefont {S.}~\bibnamefont {Iso}},\
  }\bibfield  {title} {\bibinfo {title} {{Revisiting the Naturalness Problem --
  Who is afraid of quadratic divergences? --}},\ }\href
  {https://doi.org/10.1103/PhysRevD.86.013001} {\bibfield  {journal} {\bibinfo
  {journal} {Phys. Rev. D}\ }\textbf {\bibinfo {volume} {86}},\ \bibinfo
  {pages} {013001} (\bibinfo {year} {2012})},\ \Eprint
  {https://arxiv.org/abs/1201.0857} {arXiv:1201.0857 [hep-ph]} \BibitemShut
  {NoStop}%
\bibitem [{\citenamefont {Wetterich}\ and\ \citenamefont
  {Yamada}(2017)}]{Wetterich:2016uxm}%
  \BibitemOpen
  \bibfield  {author} {\bibinfo {author} {\bibfnamefont {C.}~\bibnamefont
  {Wetterich}}\ and\ \bibinfo {author} {\bibfnamefont {M.}~\bibnamefont
  {Yamada}},\ }\bibfield  {title} {\bibinfo {title} {{Gauge hierarchy problem
  in asymptotically safe gravity--the resurgence mechanism}},\ }\href
  {https://doi.org/10.1016/j.physletb.2017.04.049} {\bibfield  {journal}
  {\bibinfo  {journal} {Phys. Lett. B}\ }\textbf {\bibinfo {volume} {770}},\
  \bibinfo {pages} {268} (\bibinfo {year} {2017})},\ \Eprint
  {https://arxiv.org/abs/1612.03069} {arXiv:1612.03069 [hep-th]} \BibitemShut
  {NoStop}%
\bibitem [{\citenamefont {Wetterich}(1988)}]{Wetterich:1987fm}%
  \BibitemOpen
  \bibfield  {author} {\bibinfo {author} {\bibfnamefont {C.}~\bibnamefont
  {Wetterich}},\ }\bibfield  {title} {\bibinfo {title} {{Cosmology and the Fate
  of Dilatation Symmetry}},\ }\href
  {https://doi.org/10.1016/0550-3213(88)90193-9} {\bibfield  {journal}
  {\bibinfo  {journal} {Nucl. Phys. B}\ }\textbf {\bibinfo {volume} {302}},\
  \bibinfo {pages} {668} (\bibinfo {year} {1988})},\ \Eprint
  {https://arxiv.org/abs/1711.03844} {arXiv:1711.03844 [hep-th]} \BibitemShut
  {NoStop}%
\bibitem [{\citenamefont {Rabinovici}\ \emph {et~al.}(1987)\citenamefont
  {Rabinovici}, \citenamefont {Saering},\ and\ \citenamefont
  {Bardeen}}]{Rabinovici:1987tf}%
  \BibitemOpen
  \bibfield  {author} {\bibinfo {author} {\bibfnamefont {E.}~\bibnamefont
  {Rabinovici}}, \bibinfo {author} {\bibfnamefont {B.}~\bibnamefont
  {Saering}},\ and\ \bibinfo {author} {\bibfnamefont {W.~A.}\ \bibnamefont
  {Bardeen}},\ }\bibfield  {title} {\bibinfo {title} {{Critical Surfaces and
  Flat Directions in a Finite Theory}},\ }\href
  {https://doi.org/10.1103/PhysRevD.36.562} {\bibfield  {journal} {\bibinfo
  {journal} {Phys. Rev. D}\ }\textbf {\bibinfo {volume} {36}},\ \bibinfo
  {pages} {562} (\bibinfo {year} {1987})}\BibitemShut {NoStop}%
\bibitem [{\citenamefont {Rubio}\ and\ \citenamefont
  {Wetterich}(2017)}]{Rubio:2017gty}%
  \BibitemOpen
  \bibfield  {author} {\bibinfo {author} {\bibfnamefont {J.}~\bibnamefont
  {Rubio}}\ and\ \bibinfo {author} {\bibfnamefont {C.}~\bibnamefont
  {Wetterich}},\ }\bibfield  {title} {\bibinfo {title} {{Emergent scale
  symmetry: Connecting inflation and dark energy}},\ }\href
  {https://doi.org/10.1103/PhysRevD.96.063509} {\bibfield  {journal} {\bibinfo
  {journal} {Phys. Rev. D}\ }\textbf {\bibinfo {volume} {96}},\ \bibinfo
  {pages} {063509} (\bibinfo {year} {2017})},\ \Eprint
  {https://arxiv.org/abs/1705.00552} {arXiv:1705.00552 [gr-qc]} \BibitemShut
  {NoStop}%
\bibitem [{\citenamefont {Rubio}(2019)}]{Rubio:2018ogq}%
  \BibitemOpen
  \bibfield  {author} {\bibinfo {author} {\bibfnamefont {J.}~\bibnamefont
  {Rubio}},\ }\bibfield  {title} {\bibinfo {title} {{Higgs inflation}},\ }\href
  {https://doi.org/10.3389/fspas.2018.00050} {\bibfield  {journal} {\bibinfo
  {journal} {Front. Astron. Space Sci.}\ }\textbf {\bibinfo {volume} {5}},\
  \bibinfo {pages} {50} (\bibinfo {year} {2019})},\ \Eprint
  {https://arxiv.org/abs/1807.02376} {arXiv:1807.02376 [hep-ph]} \BibitemShut
  {NoStop}%
\bibitem [{\citenamefont {Aghanim}\ \emph {et~al.}(2020)\citenamefont {Aghanim}
  \emph {et~al.}}]{Planck:2018vyg}%
  \BibitemOpen
  \bibfield  {author} {\bibinfo {author} {\bibfnamefont {N.}~\bibnamefont
  {Aghanim}} \emph {et~al.} (\bibinfo {collaboration} {Planck}),\ }\bibfield
  {title} {\bibinfo {title} {{Planck 2018 results. VI. Cosmological
  parameters}},\ }\href {https://doi.org/10.1051/0004-6361/201833910}
  {\bibfield  {journal} {\bibinfo  {journal} {Astron. Astrophys.}\ }\textbf
  {\bibinfo {volume} {641}},\ \bibinfo {pages} {A6} (\bibinfo {year} {2020})},\
  \bibinfo {note} {[Erratum: Astron.Astrophys. 652, C4 (2021)]},\ \Eprint
  {https://arxiv.org/abs/1807.06209} {arXiv:1807.06209 [astro-ph.CO]}
  \BibitemShut {NoStop}%
\bibitem [{\citenamefont {Akrami}\ \emph {et~al.}(2020)\citenamefont {Akrami}
  \emph {et~al.}}]{Planck:2018jri}%
  \BibitemOpen
  \bibfield  {author} {\bibinfo {author} {\bibfnamefont {Y.}~\bibnamefont
  {Akrami}} \emph {et~al.} (\bibinfo {collaboration} {Planck}),\ }\bibfield
  {title} {\bibinfo {title} {{Planck 2018 results. X. Constraints on
  inflation}},\ }\href {https://doi.org/10.1051/0004-6361/201833887} {\bibfield
   {journal} {\bibinfo  {journal} {Astron. Astrophys.}\ }\textbf {\bibinfo
  {volume} {641}},\ \bibinfo {pages} {A10} (\bibinfo {year} {2020})},\ \Eprint
  {https://arxiv.org/abs/1807.06211} {arXiv:1807.06211 [astro-ph.CO]}
  \BibitemShut {NoStop}%
\bibitem [{\citenamefont {Nakayama}(2015)}]{Nakayama:2013is}%
  \BibitemOpen
  \bibfield  {author} {\bibinfo {author} {\bibfnamefont {Y.}~\bibnamefont
  {Nakayama}},\ }\bibfield  {title} {\bibinfo {title} {{Scale invariance vs
  conformal invariance}},\ }\href
  {https://doi.org/10.1016/j.physrep.2014.12.003} {\bibfield  {journal}
  {\bibinfo  {journal} {Phys. Rept.}\ }\textbf {\bibinfo {volume} {569}},\
  \bibinfo {pages} {1} (\bibinfo {year} {2015})},\ \Eprint
  {https://arxiv.org/abs/1302.0884} {arXiv:1302.0884 [hep-th]} \BibitemShut
  {NoStop}%
\bibitem [{\citenamefont {Mannheim}(2012)}]{Mannheim:2011ds}%
  \BibitemOpen
  \bibfield  {author} {\bibinfo {author} {\bibfnamefont {P.~D.}\ \bibnamefont
  {Mannheim}},\ }\bibfield  {title} {\bibinfo {title} {{Making the Case for
  Conformal Gravity}},\ }\href {https://doi.org/10.1007/s10701-011-9608-6}
  {\bibfield  {journal} {\bibinfo  {journal} {Found. Phys.}\ }\textbf {\bibinfo
  {volume} {42}},\ \bibinfo {pages} {388} (\bibinfo {year} {2012})},\ \Eprint
  {https://arxiv.org/abs/1101.2186} {arXiv:1101.2186 [hep-th]} \BibitemShut
  {NoStop}%
\bibitem [{\citenamefont {Lucat}\ and\ \citenamefont
  {Prokopec}(2016)}]{Lucat:2016eze}%
  \BibitemOpen
  \bibfield  {author} {\bibinfo {author} {\bibfnamefont {S.}~\bibnamefont
  {Lucat}}\ and\ \bibinfo {author} {\bibfnamefont {T.}~\bibnamefont
  {Prokopec}},\ }\bibfield  {title} {\bibinfo {title} {{The role of conformal
  symmetry in gravity and the standard model}},\ }\href
  {https://doi.org/10.1088/0264-9381/33/24/245002} {\bibfield  {journal}
  {\bibinfo  {journal} {Class. Quant. Grav.}\ }\textbf {\bibinfo {volume}
  {33}},\ \bibinfo {pages} {245002} (\bibinfo {year} {2016})},\ \Eprint
  {https://arxiv.org/abs/1606.02677} {arXiv:1606.02677 [hep-th]} \BibitemShut
  {NoStop}%
\bibitem [{\citenamefont {Rachwa{\l}}(2018)}]{Rachwal:2018gwu}%
  \BibitemOpen
  \bibfield  {author} {\bibinfo {author} {\bibfnamefont {L.}~\bibnamefont
  {Rachwa{\l}}},\ }\bibfield  {title} {\bibinfo {title} {{Conformal Symmetry in
  Field Theory and in Quantum Gravity}},\ }\href
  {https://doi.org/10.3390/universe4110125} {\bibfield  {journal} {\bibinfo
  {journal} {Universe}\ }\textbf {\bibinfo {volume} {4}},\ \bibinfo {pages}
  {125} (\bibinfo {year} {2018})},\ \Eprint {https://arxiv.org/abs/1808.10457}
  {arXiv:1808.10457 [hep-th]} \BibitemShut {NoStop}%
\bibitem [{\citenamefont {Witten}(1998)}]{Witten_1998}%
  \BibitemOpen
  \bibfield  {author} {\bibinfo {author} {\bibfnamefont {E.}~\bibnamefont
  {Witten}},\ }\bibfield  {title} {\bibinfo {title} {Ads/cft correspondence and
  topological field theory},\ }\href
  {https://doi.org/10.1088/1126-6708/1998/12/012} {\bibfield  {journal}
  {\bibinfo  {journal} {Journal of High Energy Physics}\ }\textbf {\bibinfo
  {volume} {1998}},\ \bibinfo {pages} {012–012} (\bibinfo {year}
  {1998})}\BibitemShut {NoStop}%
\bibitem [{\citenamefont {Maldacena}(1999)}]{Maldacena_1999}%
  \BibitemOpen
  \bibfield  {author} {\bibinfo {author} {\bibfnamefont {J.}~\bibnamefont
  {Maldacena}},\ }\bibfield  {title} {\bibinfo {title} {The large-n limit of
  superconformal field theories and supergravity},\ }\href
  {https://doi.org/10.1023/a:1026654312961} {\bibfield  {journal} {\bibinfo
  {journal} {International Journal of Theoretical Physics}\ }\textbf {\bibinfo
  {volume} {38}},\ \bibinfo {pages} {1113–1133} (\bibinfo {year}
  {1999})}\BibitemShut {NoStop}%
\bibitem [{\citenamefont {Fitzpatrick}\ and\ \citenamefont
  {Kaplan}(2013)}]{Fitzpatrick_2013}%
  \BibitemOpen
  \bibfield  {author} {\bibinfo {author} {\bibfnamefont {A.~L.}\ \bibnamefont
  {Fitzpatrick}}\ and\ \bibinfo {author} {\bibfnamefont {J.}~\bibnamefont
  {Kaplan}},\ }\bibfield  {title} {\bibinfo {title} {Ads field theory from
  conformal field theory},\ }\bibfield  {journal} {\bibinfo  {journal} {Journal
  of High Energy Physics}\ }\textbf {\bibinfo {volume} {2013}},\ \href
  {https://doi.org/10.1007/jhep02(2013)054} {10.1007/jhep02(2013)054} (\bibinfo
  {year} {2013})\BibitemShut {NoStop}%
\bibitem [{\citenamefont {Chamseddine}\ and\ \citenamefont
  {Mukhanov}(2013{\natexlab{a}})}]{Chamseddine:2013kea}%
  \BibitemOpen
  \bibfield  {author} {\bibinfo {author} {\bibfnamefont {A.~H.}\ \bibnamefont
  {Chamseddine}}\ and\ \bibinfo {author} {\bibfnamefont {V.}~\bibnamefont
  {Mukhanov}},\ }\bibfield  {title} {\bibinfo {title} {{Mimetic Dark Matter}},\
  }\href {https://doi.org/10.1007/JHEP11(2013)135} {\bibfield  {journal}
  {\bibinfo  {journal} {JHEP}\ }\textbf {\bibinfo {volume} {11}},\ \bibinfo
  {pages} {135}},\ \Eprint {https://arxiv.org/abs/1308.5410} {arXiv:1308.5410
  [astro-ph.CO]} \BibitemShut {NoStop}%
\bibitem [{\citenamefont {Golovnev}(2014)}]{Golovnev:2013jxa}%
  \BibitemOpen
  \bibfield  {author} {\bibinfo {author} {\bibfnamefont {A.}~\bibnamefont
  {Golovnev}},\ }\bibfield  {title} {\bibinfo {title} {{On the recently
  proposed Mimetic Dark Matter}},\ }\href
  {https://doi.org/10.1016/j.physletb.2013.11.026} {\bibfield  {journal}
  {\bibinfo  {journal} {Phys. Lett. B}\ }\textbf {\bibinfo {volume} {728}},\
  \bibinfo {pages} {39} (\bibinfo {year} {2014})},\ \Eprint
  {https://arxiv.org/abs/1310.2790} {arXiv:1310.2790 [gr-qc]} \BibitemShut
  {NoStop}%
\bibitem [{\citenamefont {Barvinsky}(2014)}]{Barvinsky:2013mea}%
  \BibitemOpen
  \bibfield  {author} {\bibinfo {author} {\bibfnamefont {A.~O.}\ \bibnamefont
  {Barvinsky}},\ }\bibfield  {title} {\bibinfo {title} {{Dark matter as a ghost
  free conformal extension of Einstein theory}},\ }\href
  {https://doi.org/10.1088/1475-7516/2014/01/014} {\bibfield  {journal}
  {\bibinfo  {journal} {JCAP}\ }\textbf {\bibinfo {volume} {01}},\ \bibinfo
  {pages} {014}},\ \Eprint {https://arxiv.org/abs/1311.3111} {arXiv:1311.3111
  [hep-th]} \BibitemShut {NoStop}%
\bibitem [{\citenamefont {Deruelle}\ and\ \citenamefont
  {Rua}(2014)}]{Deruelle:2014zza}%
  \BibitemOpen
  \bibfield  {author} {\bibinfo {author} {\bibfnamefont {N.}~\bibnamefont
  {Deruelle}}\ and\ \bibinfo {author} {\bibfnamefont {J.}~\bibnamefont {Rua}},\
  }\bibfield  {title} {\bibinfo {title} {{Disformal Transformations, Veiled
  General Relativity and Mimetic Gravity}},\ }\href
  {https://doi.org/10.1088/1475-7516/2014/09/002} {\bibfield  {journal}
  {\bibinfo  {journal} {JCAP}\ }\textbf {\bibinfo {volume} {09}},\ \bibinfo
  {pages} {002}},\ \Eprint {https://arxiv.org/abs/1407.0825} {arXiv:1407.0825
  [gr-qc]} \BibitemShut {NoStop}%
\bibitem [{\citenamefont {Arroja}\ \emph {et~al.}(2015)\citenamefont {Arroja},
  \citenamefont {Bartolo}, \citenamefont {Karmakar},\ and\ \citenamefont
  {Matarrese}}]{Arroja:2015wpa}%
  \BibitemOpen
  \bibfield  {author} {\bibinfo {author} {\bibfnamefont {F.}~\bibnamefont
  {Arroja}}, \bibinfo {author} {\bibfnamefont {N.}~\bibnamefont {Bartolo}},
  \bibinfo {author} {\bibfnamefont {P.}~\bibnamefont {Karmakar}},\ and\
  \bibinfo {author} {\bibfnamefont {S.}~\bibnamefont {Matarrese}},\ }\bibfield
  {title} {\bibinfo {title} {{The two faces of mimetic Horndeski gravity:
  disformal transformations and Lagrange multiplier}},\ }\href
  {https://doi.org/10.1088/1475-7516/2015/09/051} {\bibfield  {journal}
  {\bibinfo  {journal} {JCAP}\ }\textbf {\bibinfo {volume} {09}},\ \bibinfo
  {pages} {051}},\ \Eprint {https://arxiv.org/abs/1506.08575} {arXiv:1506.08575
  [gr-qc]} \BibitemShut {NoStop}%
\bibitem [{\citenamefont {Jirou{\v{s}}ek}\ \emph {et~al.}(2022)\citenamefont
  {Jirou{\v{s}}ek}, \citenamefont {Shimada}, \citenamefont {Vikman},\ and\
  \citenamefont {Yamaguchi}}]{Jirousek:2022rym}%
  \BibitemOpen
  \bibfield  {author} {\bibinfo {author} {\bibfnamefont {P.}~\bibnamefont
  {Jirou{\v{s}}ek}}, \bibinfo {author} {\bibfnamefont {K.}~\bibnamefont
  {Shimada}}, \bibinfo {author} {\bibfnamefont {A.}~\bibnamefont {Vikman}},\
  and\ \bibinfo {author} {\bibfnamefont {M.}~\bibnamefont {Yamaguchi}},\
  }\bibfield  {title} {\bibinfo {title} {{Disforming to conformal symmetry}},\
  }\href {https://doi.org/10.1088/1475-7516/2022/11/019} {\bibfield  {journal}
  {\bibinfo  {journal} {JCAP}\ }\textbf {\bibinfo {volume} {11}},\ \bibinfo
  {pages} {019}},\ \Eprint {https://arxiv.org/abs/2207.12611} {arXiv:2207.12611
  [gr-qc]} \BibitemShut {NoStop}%
\bibitem [{\citenamefont {Golovnev}(2023)}]{Golovnev:2022jts}%
  \BibitemOpen
  \bibfield  {author} {\bibinfo {author} {\bibfnamefont {A.}~\bibnamefont
  {Golovnev}},\ }\bibfield  {title} {\bibinfo {title} {{The variational
  principle, conformal and disformal transformations, and the degrees of
  freedom}},\ }\href {https://doi.org/10.1063/5.0120079} {\bibfield  {journal}
  {\bibinfo  {journal} {J. Math. Phys.}\ }\textbf {\bibinfo {volume} {64}},\
  \bibinfo {pages} {012501} (\bibinfo {year} {2023})},\ \Eprint
  {https://arxiv.org/abs/2208.04082} {arXiv:2208.04082 [gr-qc]} \BibitemShut
  {NoStop}%
\bibitem [{\citenamefont {Jirou{\v{s}}ek}\ \emph {et~al.}(2023)\citenamefont
  {Jirou{\v{s}}ek}, \citenamefont {Shimada}, \citenamefont {Vikman},\ and\
  \citenamefont {Yamaguchi}}]{Jirousek:2022jhh}%
  \BibitemOpen
  \bibfield  {author} {\bibinfo {author} {\bibfnamefont {P.}~\bibnamefont
  {Jirou{\v{s}}ek}}, \bibinfo {author} {\bibfnamefont {K.}~\bibnamefont
  {Shimada}}, \bibinfo {author} {\bibfnamefont {A.}~\bibnamefont {Vikman}},\
  and\ \bibinfo {author} {\bibfnamefont {M.}~\bibnamefont {Yamaguchi}},\
  }\bibfield  {title} {\bibinfo {title} {{New dynamical degrees of freedom from
  invertible transformations}},\ }\href
  {https://doi.org/10.1007/JHEP07(2023)154} {\bibfield  {journal} {\bibinfo
  {journal} {JHEP}\ }\textbf {\bibinfo {volume} {07}},\ \bibinfo {pages}
  {154}},\ \Eprint {https://arxiv.org/abs/2208.05951} {arXiv:2208.05951
  [gr-qc]} \BibitemShut {NoStop}%
\bibitem [{\citenamefont {Hammer}\ and\ \citenamefont
  {Vikman}(2015)}]{Hammer:2015pcx}%
  \BibitemOpen
  \bibfield  {author} {\bibinfo {author} {\bibfnamefont {K.}~\bibnamefont
  {Hammer}}\ and\ \bibinfo {author} {\bibfnamefont {A.}~\bibnamefont
  {Vikman}},\ }\bibfield  {title} {\bibinfo {title} {{Many Faces of Mimetic
  Gravity}},\ }\href@noop {} {\  (\bibinfo {year} {2015})},\ \Eprint
  {https://arxiv.org/abs/1512.09118} {arXiv:1512.09118 [gr-qc]} \BibitemShut
  {NoStop}%
\bibitem [{\citenamefont {Babichev}\ \emph {et~al.}(2024)\citenamefont
  {Babichev}, \citenamefont {Izumi}, \citenamefont {Noui}, \citenamefont
  {Tanahashi},\ and\ \citenamefont {Yamaguchi}}]{Babichev:2024eoh}%
  \BibitemOpen
  \bibfield  {author} {\bibinfo {author} {\bibfnamefont {E.}~\bibnamefont
  {Babichev}}, \bibinfo {author} {\bibfnamefont {K.}~\bibnamefont {Izumi}},
  \bibinfo {author} {\bibfnamefont {K.}~\bibnamefont {Noui}}, \bibinfo {author}
  {\bibfnamefont {N.}~\bibnamefont {Tanahashi}},\ and\ \bibinfo {author}
  {\bibfnamefont {M.}~\bibnamefont {Yamaguchi}},\ }\bibfield  {title} {\bibinfo
  {title} {{Generalization of conformal-disformal transformations of the metric
  in scalar-tensor theories}},\ }\href
  {https://doi.org/10.1103/PhysRevD.110.064063} {\bibfield  {journal} {\bibinfo
   {journal} {Phys. Rev. D}\ }\textbf {\bibinfo {volume} {110}},\ \bibinfo
  {pages} {064063} (\bibinfo {year} {2024})},\ \Eprint
  {https://arxiv.org/abs/2405.13126} {arXiv:2405.13126 [gr-qc]} \BibitemShut
  {NoStop}%
\bibitem [{\citenamefont {Dom{\`e}nech}\ and\ \citenamefont
  {Ganz}(2023)}]{Domenech:2023ryc}%
  \BibitemOpen
  \bibfield  {author} {\bibinfo {author} {\bibfnamefont {G.}~\bibnamefont
  {Dom{\`e}nech}}\ and\ \bibinfo {author} {\bibfnamefont {A.}~\bibnamefont
  {Ganz}},\ }\bibfield  {title} {\bibinfo {title} {{Disformal symmetry in the
  Universe: mimetic gravity and beyond}},\ }\href
  {https://doi.org/10.1088/1475-7516/2023/08/046} {\bibfield  {journal}
  {\bibinfo  {journal} {JCAP}\ }\textbf {\bibinfo {volume} {08}},\ \bibinfo
  {pages} {046}},\ \Eprint {https://arxiv.org/abs/2304.11035} {arXiv:2304.11035
  [gr-qc]} \BibitemShut {NoStop}%
\bibitem [{\citenamefont {Dom{\`e}nech}\ and\ \citenamefont
  {Ganz}(2025)}]{Domenech:2025qny}%
  \BibitemOpen
  \bibfield  {author} {\bibinfo {author} {\bibfnamefont {G.}~\bibnamefont
  {Dom{\`e}nech}}\ and\ \bibinfo {author} {\bibfnamefont {A.}~\bibnamefont
  {Ganz}},\ }\bibfield  {title} {\bibinfo {title} {{Connecting relativistic
  MOND theories with mimetic gravity}},\ }\href
  {https://doi.org/10.1088/1475-7516/2025/06/059} {\bibfield  {journal}
  {\bibinfo  {journal} {JCAP}\ }\textbf {\bibinfo {volume} {06}},\ \bibinfo
  {pages} {059}},\ \Eprint {https://arxiv.org/abs/2503.11174} {arXiv:2503.11174
  [gr-qc]} \BibitemShut {NoStop}%
\bibitem [{\citenamefont {Sebastiani}\ \emph {et~al.}(2017)\citenamefont
  {Sebastiani}, \citenamefont {Vagnozzi},\ and\ \citenamefont
  {Myrzakulov}}]{Sebastiani:2016ras}%
  \BibitemOpen
  \bibfield  {author} {\bibinfo {author} {\bibfnamefont {L.}~\bibnamefont
  {Sebastiani}}, \bibinfo {author} {\bibfnamefont {S.}~\bibnamefont
  {Vagnozzi}},\ and\ \bibinfo {author} {\bibfnamefont {R.}~\bibnamefont
  {Myrzakulov}},\ }\bibfield  {title} {\bibinfo {title} {{Mimetic gravity: a
  review of recent developments and applications to cosmology and
  astrophysics}},\ }\href {https://doi.org/10.1155/2017/3156915} {\bibfield
  {journal} {\bibinfo  {journal} {Adv. High Energy Phys.}\ }\textbf {\bibinfo
  {volume} {2017}},\ \bibinfo {pages} {3156915} (\bibinfo {year} {2017})},\
  \Eprint {https://arxiv.org/abs/1612.08661} {arXiv:1612.08661 [gr-qc]}
  \BibitemShut {NoStop}%
\bibitem [{\citenamefont {Malaeb}(2026)}]{Malaeb:2026ecv}%
  \BibitemOpen
  \bibfield  {author} {\bibinfo {author} {\bibfnamefont {O.}~\bibnamefont
  {Malaeb}},\ }\bibfield  {title} {\bibinfo {title} {{The Theoretical Landscape
  of Mimetic Gravity: A Comprehensive Review}},\ }\href@noop {} {\  (\bibinfo
  {year} {2026})},\ \Eprint {https://arxiv.org/abs/2602.14082}
  {arXiv:2602.14082 [gr-qc]} \BibitemShut {NoStop}%
\bibitem [{\citenamefont {Boyle}\ and\ \citenamefont
  {Turok}(2021)}]{Boyle:2021jaz}%
  \BibitemOpen
  \bibfield  {author} {\bibinfo {author} {\bibfnamefont {L.}~\bibnamefont
  {Boyle}}\ and\ \bibinfo {author} {\bibfnamefont {N.}~\bibnamefont {Turok}},\
  }\bibfield  {title} {\bibinfo {title} {{Cancelling the vacuum energy and Weyl
  anomaly in the standard model with dimension-zero scalar fields}},\
  }\href@noop {} {\  (\bibinfo {year} {2021})},\ \Eprint
  {https://arxiv.org/abs/2110.06258} {arXiv:2110.06258 [hep-th]} \BibitemShut
  {NoStop}%
\bibitem [{\citenamefont {Turok}\ and\ \citenamefont
  {Boyle}(2023)}]{Turok:2023amx}%
  \BibitemOpen
  \bibfield  {author} {\bibinfo {author} {\bibfnamefont {N.}~\bibnamefont
  {Turok}}\ and\ \bibinfo {author} {\bibfnamefont {L.}~\bibnamefont {Boyle}},\
  }\bibfield  {title} {\bibinfo {title} {{A Minimal Explanation of the
  Primordial Cosmological Perturbations}},\ }\href@noop {} {\  (\bibinfo {year}
  {2023})},\ \Eprint {https://arxiv.org/abs/2302.00344} {arXiv:2302.00344
  [hep-ph]} \BibitemShut {NoStop}%
\bibitem [{\citenamefont {Cline}\ and\ \citenamefont
  {Hell}(2026)}]{Cline:2026zcv}%
  \BibitemOpen
  \bibfield  {author} {\bibinfo {author} {\bibfnamefont {J.~M.}\ \bibnamefont
  {Cline}}\ and\ \bibinfo {author} {\bibfnamefont {A.}~\bibnamefont {Hell}},\
  }\bibfield  {title} {\bibinfo {title} {{Pathologies of dimension-zero scalar
  fields}},\ }\href@noop {} {\  (\bibinfo {year} {2026})},\ \Eprint
  {https://arxiv.org/abs/2603.05683} {arXiv:2603.05683 [hep-th]} \BibitemShut
  {NoStop}%
\bibitem [{\citenamefont {Takahashi}\ and\ \citenamefont
  {Kobayashi}(2017)}]{Takahashi:2017pje}%
  \BibitemOpen
  \bibfield  {author} {\bibinfo {author} {\bibfnamefont {K.}~\bibnamefont
  {Takahashi}}\ and\ \bibinfo {author} {\bibfnamefont {T.}~\bibnamefont
  {Kobayashi}},\ }\bibfield  {title} {\bibinfo {title} {{Extended mimetic
  gravity: Hamiltonian analysis and gradient instabilities}},\ }\href
  {https://doi.org/10.1088/1475-7516/2017/11/038} {\bibfield  {journal}
  {\bibinfo  {journal} {JCAP}\ }\textbf {\bibinfo {volume} {11}},\ \bibinfo
  {pages} {038}},\ \Eprint {https://arxiv.org/abs/1708.02951} {arXiv:1708.02951
  [gr-qc]} \BibitemShut {NoStop}%
\bibitem [{\citenamefont {Chamseddine}\ \emph {et~al.}(2014)\citenamefont
  {Chamseddine}, \citenamefont {Mukhanov},\ and\ \citenamefont
  {Vikman}}]{Chamseddine:2014vna}%
  \BibitemOpen
  \bibfield  {author} {\bibinfo {author} {\bibfnamefont {A.~H.}\ \bibnamefont
  {Chamseddine}}, \bibinfo {author} {\bibfnamefont {V.}~\bibnamefont
  {Mukhanov}},\ and\ \bibinfo {author} {\bibfnamefont {A.}~\bibnamefont
  {Vikman}},\ }\bibfield  {title} {\bibinfo {title} {{Cosmology with Mimetic
  Matter}},\ }\href {https://doi.org/10.1088/1475-7516/2014/06/017} {\bibfield
  {journal} {\bibinfo  {journal} {JCAP}\ }\textbf {\bibinfo {volume} {06}},\
  \bibinfo {pages} {017}},\ \Eprint {https://arxiv.org/abs/1403.3961}
  {arXiv:1403.3961 [astro-ph.CO]} \BibitemShut {NoStop}%
\bibitem [{\citenamefont {Chaichian}\ \emph {et~al.}(2014)\citenamefont
  {Chaichian}, \citenamefont {Kluson}, \citenamefont {Oksanen},\ and\
  \citenamefont {Tureanu}}]{Chaichian:2014qba}%
  \BibitemOpen
  \bibfield  {author} {\bibinfo {author} {\bibfnamefont {M.}~\bibnamefont
  {Chaichian}}, \bibinfo {author} {\bibfnamefont {J.}~\bibnamefont {Kluson}},
  \bibinfo {author} {\bibfnamefont {M.}~\bibnamefont {Oksanen}},\ and\ \bibinfo
  {author} {\bibfnamefont {A.}~\bibnamefont {Tureanu}},\ }\bibfield  {title}
  {\bibinfo {title} {{Mimetic dark matter, ghost instability and a mimetic
  tensor-vector-scalar gravity}},\ }\href
  {https://doi.org/10.1007/JHEP12(2014)102} {\bibfield  {journal} {\bibinfo
  {journal} {JHEP}\ }\textbf {\bibinfo {volume} {12}},\ \bibinfo {pages}
  {102}},\ \Eprint {https://arxiv.org/abs/1404.4008} {arXiv:1404.4008 [hep-th]}
  \BibitemShut {NoStop}%
\bibitem [{\citenamefont {Mirzagholi}\ and\ \citenamefont
  {Vikman}(2015)}]{Mirzagholi:2014ifa}%
  \BibitemOpen
  \bibfield  {author} {\bibinfo {author} {\bibfnamefont {L.}~\bibnamefont
  {Mirzagholi}}\ and\ \bibinfo {author} {\bibfnamefont {A.}~\bibnamefont
  {Vikman}},\ }\bibfield  {title} {\bibinfo {title} {{Imperfect Dark Matter}},\
  }\href {https://doi.org/10.1088/1475-7516/2015/06/028} {\bibfield  {journal}
  {\bibinfo  {journal} {JCAP}\ }\textbf {\bibinfo {volume} {06}},\ \bibinfo
  {pages} {028}},\ \Eprint {https://arxiv.org/abs/1412.7136} {arXiv:1412.7136
  [gr-qc]} \BibitemShut {NoStop}%
\bibitem [{\citenamefont {Langlois}\ \emph {et~al.}(2019)\citenamefont
  {Langlois}, \citenamefont {Mancarella}, \citenamefont {Noui},\ and\
  \citenamefont {Vernizzi}}]{Langlois:2018jdg}%
  \BibitemOpen
  \bibfield  {author} {\bibinfo {author} {\bibfnamefont {D.}~\bibnamefont
  {Langlois}}, \bibinfo {author} {\bibfnamefont {M.}~\bibnamefont
  {Mancarella}}, \bibinfo {author} {\bibfnamefont {K.}~\bibnamefont {Noui}},\
  and\ \bibinfo {author} {\bibfnamefont {F.}~\bibnamefont {Vernizzi}},\
  }\bibfield  {title} {\bibinfo {title} {{Mimetic gravity as DHOST theories}},\
  }\href {https://doi.org/10.1088/1475-7516/2019/02/036} {\bibfield  {journal}
  {\bibinfo  {journal} {JCAP}\ }\textbf {\bibinfo {volume} {02}},\ \bibinfo
  {pages} {036}},\ \Eprint {https://arxiv.org/abs/1802.03394} {arXiv:1802.03394
  [gr-qc]} \BibitemShut {NoStop}%
\bibitem [{\citenamefont {Langlois}\ and\ \citenamefont
  {Noui}(2016)}]{Langlois:2015cwa}%
  \BibitemOpen
  \bibfield  {author} {\bibinfo {author} {\bibfnamefont {D.}~\bibnamefont
  {Langlois}}\ and\ \bibinfo {author} {\bibfnamefont {K.}~\bibnamefont
  {Noui}},\ }\bibfield  {title} {\bibinfo {title} {{Degenerate higher
  derivative theories beyond Horndeski: evading the Ostrogradski
  instability}},\ }\href {https://doi.org/10.1088/1475-7516/2016/02/034}
  {\bibfield  {journal} {\bibinfo  {journal} {JCAP}\ }\textbf {\bibinfo
  {volume} {02}},\ \bibinfo {pages} {034}},\ \Eprint
  {https://arxiv.org/abs/1510.06930} {arXiv:1510.06930 [gr-qc]} \BibitemShut
  {NoStop}%
\bibitem [{\citenamefont {Ben~Achour}\ \emph
  {et~al.}(2016{\natexlab{a}})\citenamefont {Ben~Achour}, \citenamefont
  {Langlois},\ and\ \citenamefont {Noui}}]{BenAchour:2016cay}%
  \BibitemOpen
  \bibfield  {author} {\bibinfo {author} {\bibfnamefont {J.}~\bibnamefont
  {Ben~Achour}}, \bibinfo {author} {\bibfnamefont {D.}~\bibnamefont
  {Langlois}},\ and\ \bibinfo {author} {\bibfnamefont {K.}~\bibnamefont
  {Noui}},\ }\bibfield  {title} {\bibinfo {title} {{Degenerate higher order
  scalar-tensor theories beyond Horndeski and disformal transformations}},\
  }\href {https://doi.org/10.1103/PhysRevD.93.124005} {\bibfield  {journal}
  {\bibinfo  {journal} {Phys. Rev. D}\ }\textbf {\bibinfo {volume} {93}},\
  \bibinfo {pages} {124005} (\bibinfo {year} {2016}{\natexlab{a}})},\ \Eprint
  {https://arxiv.org/abs/1602.08398} {arXiv:1602.08398 [gr-qc]} \BibitemShut
  {NoStop}%
\bibitem [{\citenamefont {Ben~Achour}\ \emph
  {et~al.}(2016{\natexlab{b}})\citenamefont {Ben~Achour}, \citenamefont
  {Crisostomi}, \citenamefont {Koyama}, \citenamefont {Langlois}, \citenamefont
  {Noui},\ and\ \citenamefont {Tasinato}}]{BenAchour:2016fzp}%
  \BibitemOpen
  \bibfield  {author} {\bibinfo {author} {\bibfnamefont {J.}~\bibnamefont
  {Ben~Achour}}, \bibinfo {author} {\bibfnamefont {M.}~\bibnamefont
  {Crisostomi}}, \bibinfo {author} {\bibfnamefont {K.}~\bibnamefont {Koyama}},
  \bibinfo {author} {\bibfnamefont {D.}~\bibnamefont {Langlois}}, \bibinfo
  {author} {\bibfnamefont {K.}~\bibnamefont {Noui}},\ and\ \bibinfo {author}
  {\bibfnamefont {G.}~\bibnamefont {Tasinato}},\ }\bibfield  {title} {\bibinfo
  {title} {{Degenerate higher order scalar-tensor theories beyond Horndeski up
  to cubic order}},\ }\href {https://doi.org/10.1007/JHEP12(2016)100}
  {\bibfield  {journal} {\bibinfo  {journal} {JHEP}\ }\textbf {\bibinfo
  {volume} {12}},\ \bibinfo {pages} {100}},\ \Eprint
  {https://arxiv.org/abs/1608.08135} {arXiv:1608.08135 [hep-th]} \BibitemShut
  {NoStop}%
\bibitem [{\citenamefont {Crisostomi}\ \emph {et~al.}(2016)\citenamefont
  {Crisostomi}, \citenamefont {Koyama},\ and\ \citenamefont
  {Tasinato}}]{Crisostomi:2016czh}%
  \BibitemOpen
  \bibfield  {author} {\bibinfo {author} {\bibfnamefont {M.}~\bibnamefont
  {Crisostomi}}, \bibinfo {author} {\bibfnamefont {K.}~\bibnamefont {Koyama}},\
  and\ \bibinfo {author} {\bibfnamefont {G.}~\bibnamefont {Tasinato}},\
  }\bibfield  {title} {\bibinfo {title} {{Extended Scalar-Tensor Theories of
  Gravity}},\ }\href {https://doi.org/10.1088/1475-7516/2016/04/044} {\bibfield
   {journal} {\bibinfo  {journal} {JCAP}\ }\textbf {\bibinfo {volume} {04}},\
  \bibinfo {pages} {044}},\ \Eprint {https://arxiv.org/abs/1602.03119}
  {arXiv:1602.03119 [hep-th]} \BibitemShut {NoStop}%
\bibitem [{\citenamefont {de~Rham}\ and\ \citenamefont
  {Matas}(2016)}]{deRham:2016wji}%
  \BibitemOpen
  \bibfield  {author} {\bibinfo {author} {\bibfnamefont {C.}~\bibnamefont
  {de~Rham}}\ and\ \bibinfo {author} {\bibfnamefont {A.}~\bibnamefont
  {Matas}},\ }\bibfield  {title} {\bibinfo {title} {{Ostrogradsky in Theories
  with Multiple Fields}},\ }\href
  {https://doi.org/10.1088/1475-7516/2016/06/041} {\bibfield  {journal}
  {\bibinfo  {journal} {JCAP}\ }\textbf {\bibinfo {volume} {06}},\ \bibinfo
  {pages} {041}},\ \Eprint {https://arxiv.org/abs/1604.08638} {arXiv:1604.08638
  [hep-th]} \BibitemShut {NoStop}%
\bibitem [{\citenamefont {Ganz}\ \emph
  {et~al.}(2019{\natexlab{a}})\citenamefont {Ganz}, \citenamefont {Karmakar},
  \citenamefont {Matarrese},\ and\ \citenamefont {Sorokin}}]{Ganz:2018mqi}%
  \BibitemOpen
  \bibfield  {author} {\bibinfo {author} {\bibfnamefont {A.}~\bibnamefont
  {Ganz}}, \bibinfo {author} {\bibfnamefont {P.}~\bibnamefont {Karmakar}},
  \bibinfo {author} {\bibfnamefont {S.}~\bibnamefont {Matarrese}},\ and\
  \bibinfo {author} {\bibfnamefont {D.}~\bibnamefont {Sorokin}},\ }\bibfield
  {title} {\bibinfo {title} {{Hamiltonian analysis of mimetic scalar gravity
  revisited}},\ }\href {https://doi.org/10.1103/PhysRevD.99.064009} {\bibfield
  {journal} {\bibinfo  {journal} {Phys. Rev. D}\ }\textbf {\bibinfo {volume}
  {99}},\ \bibinfo {pages} {064009} (\bibinfo {year} {2019}{\natexlab{a}})},\
  \Eprint {https://arxiv.org/abs/1812.02667} {arXiv:1812.02667 [gr-qc]}
  \BibitemShut {NoStop}%
\bibitem [{\citenamefont {Ganz}\ \emph
  {et~al.}(2019{\natexlab{b}})\citenamefont {Ganz}, \citenamefont {Bartolo},\
  and\ \citenamefont {Matarrese}}]{Ganz:2019vre}%
  \BibitemOpen
  \bibfield  {author} {\bibinfo {author} {\bibfnamefont {A.}~\bibnamefont
  {Ganz}}, \bibinfo {author} {\bibfnamefont {N.}~\bibnamefont {Bartolo}},\ and\
  \bibinfo {author} {\bibfnamefont {S.}~\bibnamefont {Matarrese}},\ }\bibfield
  {title} {\bibinfo {title} {{Towards a viable effective field theory of
  mimetic gravity}},\ }\href {https://doi.org/10.1088/1475-7516/2019/12/037}
  {\bibfield  {journal} {\bibinfo  {journal} {JCAP}\ }\textbf {\bibinfo
  {volume} {12}},\ \bibinfo {pages} {037}},\ \Eprint
  {https://arxiv.org/abs/1907.10301} {arXiv:1907.10301 [gr-qc]} \BibitemShut
  {NoStop}%
\bibitem [{\citenamefont {Firouzjahi}\ \emph {et~al.}(2017)\citenamefont
  {Firouzjahi}, \citenamefont {Gorji},\ and\ \citenamefont
  {Hosseini~Mansoori}}]{Firouzjahi:2017txv}%
  \BibitemOpen
  \bibfield  {author} {\bibinfo {author} {\bibfnamefont {H.}~\bibnamefont
  {Firouzjahi}}, \bibinfo {author} {\bibfnamefont {M.~A.}\ \bibnamefont
  {Gorji}},\ and\ \bibinfo {author} {\bibfnamefont {S.~A.}\ \bibnamefont
  {Hosseini~Mansoori}},\ }\bibfield  {title} {\bibinfo {title} {{Instabilities
  in Mimetic Matter Perturbations}},\ }\href
  {https://doi.org/10.1088/1475-7516/2017/07/031} {\bibfield  {journal}
  {\bibinfo  {journal} {JCAP}\ }\textbf {\bibinfo {volume} {07}},\ \bibinfo
  {pages} {031}},\ \Eprint {https://arxiv.org/abs/1703.02923} {arXiv:1703.02923
  [hep-th]} \BibitemShut {NoStop}%
\bibitem [{\citenamefont {Hirano}\ \emph {et~al.}(2017)\citenamefont {Hirano},
  \citenamefont {Nishi},\ and\ \citenamefont {Kobayashi}}]{Hirano:2017zox}%
  \BibitemOpen
  \bibfield  {author} {\bibinfo {author} {\bibfnamefont {S.}~\bibnamefont
  {Hirano}}, \bibinfo {author} {\bibfnamefont {S.}~\bibnamefont {Nishi}},\ and\
  \bibinfo {author} {\bibfnamefont {T.}~\bibnamefont {Kobayashi}},\ }\bibfield
  {title} {\bibinfo {title} {{Healthy imperfect dark matter from effective
  theory of mimetic cosmological perturbations}},\ }\href
  {https://doi.org/10.1088/1475-7516/2017/07/009} {\bibfield  {journal}
  {\bibinfo  {journal} {JCAP}\ }\textbf {\bibinfo {volume} {07}},\ \bibinfo
  {pages} {009}},\ \Eprint {https://arxiv.org/abs/1704.06031} {arXiv:1704.06031
  [gr-qc]} \BibitemShut {NoStop}%
\bibitem [{\citenamefont {Zheng}\ \emph {et~al.}(2017)\citenamefont {Zheng},
  \citenamefont {Shen}, \citenamefont {Mou},\ and\ \citenamefont
  {Li}}]{Zheng:2017qfs}%
  \BibitemOpen
  \bibfield  {author} {\bibinfo {author} {\bibfnamefont {Y.}~\bibnamefont
  {Zheng}}, \bibinfo {author} {\bibfnamefont {L.}~\bibnamefont {Shen}},
  \bibinfo {author} {\bibfnamefont {Y.}~\bibnamefont {Mou}},\ and\ \bibinfo
  {author} {\bibfnamefont {M.}~\bibnamefont {Li}},\ }\bibfield  {title}
  {\bibinfo {title} {{On (in)stabilities of perturbations in mimetic models
  with higher derivatives}},\ }\href
  {https://doi.org/10.1088/1475-7516/2017/08/040} {\bibfield  {journal}
  {\bibinfo  {journal} {JCAP}\ }\textbf {\bibinfo {volume} {08}},\ \bibinfo
  {pages} {040}},\ \Eprint {https://arxiv.org/abs/1704.06834} {arXiv:1704.06834
  [gr-qc]} \BibitemShut {NoStop}%
\bibitem [{\citenamefont {Gorji}\ \emph {et~al.}(2018)\citenamefont {Gorji},
  \citenamefont {Hosseini~Mansoori},\ and\ \citenamefont
  {Firouzjahi}}]{Gorji:2017cai}%
  \BibitemOpen
  \bibfield  {author} {\bibinfo {author} {\bibfnamefont {M.~A.}\ \bibnamefont
  {Gorji}}, \bibinfo {author} {\bibfnamefont {S.~A.}\ \bibnamefont
  {Hosseini~Mansoori}},\ and\ \bibinfo {author} {\bibfnamefont
  {H.}~\bibnamefont {Firouzjahi}},\ }\bibfield  {title} {\bibinfo {title}
  {{Higher Derivative Mimetic Gravity}},\ }\href
  {https://doi.org/10.1088/1475-7516/2018/01/020} {\bibfield  {journal}
  {\bibinfo  {journal} {JCAP}\ }\textbf {\bibinfo {volume} {01}},\ \bibinfo
  {pages} {020}},\ \Eprint {https://arxiv.org/abs/1709.09988} {arXiv:1709.09988
  [astro-ph.CO]} \BibitemShut {NoStop}%
\bibitem [{\citenamefont {Abbott}\ \emph {et~al.}(2017)\citenamefont {Abbott}
  \emph {et~al.}}]{LIGOScientific:2017zic}%
  \BibitemOpen
  \bibfield  {author} {\bibinfo {author} {\bibfnamefont {B.~P.}\ \bibnamefont
  {Abbott}} \emph {et~al.} (\bibinfo {collaboration} {LIGO Scientific, Virgo,
  Fermi-GBM, INTEGRAL}),\ }\bibfield  {title} {\bibinfo {title} {{Gravitational
  Waves and Gamma-rays from a Binary Neutron Star Merger: GW170817 and GRB
  170817A}},\ }\href {https://doi.org/10.3847/2041-8213/aa920c} {\bibfield
  {journal} {\bibinfo  {journal} {Astrophys. J. Lett.}\ }\textbf {\bibinfo
  {volume} {848}},\ \bibinfo {pages} {L13} (\bibinfo {year} {2017})},\ \Eprint
  {https://arxiv.org/abs/1710.05834} {arXiv:1710.05834 [astro-ph.HE]}
  \BibitemShut {NoStop}%
\bibitem [{\citenamefont {Paneitz}(2008)}]{Paneitz:2008afy}%
  \BibitemOpen
  \bibfield  {author} {\bibinfo {author} {\bibfnamefont {S.~M.}\ \bibnamefont
  {Paneitz}},\ }\bibfield  {title} {\bibinfo {title} {{A Quartic Conformally
  Covariant Differential Operator for Arbitrary Pseudo-Riemannian Manifolds
  (Summary)}}\ }\href {https://doi.org/10.3842/sigma.2008.036}
  {10.3842/sigma.2008.036} (\bibinfo {year} {2008}),\ \Eprint
  {https://arxiv.org/abs/0803.4331} {arXiv:0803.4331 [math.DG]} \BibitemShut
  {NoStop}%
\bibitem [{\citenamefont {Kaku}\ \emph {et~al.}(1978)\citenamefont {Kaku},
  \citenamefont {Townsend},\ and\ \citenamefont {van
  Nieuwenhuizen}}]{Kaku:1978nz}%
  \BibitemOpen
  \bibfield  {author} {\bibinfo {author} {\bibfnamefont {M.}~\bibnamefont
  {Kaku}}, \bibinfo {author} {\bibfnamefont {P.~K.}\ \bibnamefont {Townsend}},\
  and\ \bibinfo {author} {\bibfnamefont {P.}~\bibnamefont {van
  Nieuwenhuizen}},\ }\bibfield  {title} {\bibinfo {title} {{Properties of
  Conformal Supergravity}},\ }\href {https://doi.org/10.1103/PhysRevD.17.3179}
  {\bibfield  {journal} {\bibinfo  {journal} {Phys. Rev. D}\ }\textbf {\bibinfo
  {volume} {17}},\ \bibinfo {pages} {3179} (\bibinfo {year}
  {1978})}\BibitemShut {NoStop}%
\bibitem [{\citenamefont {Fradkin}\ and\ \citenamefont
  {Tseytlin}(1982{\natexlab{a}})}]{Fradkin:1981jc}%
  \BibitemOpen
  \bibfield  {author} {\bibinfo {author} {\bibfnamefont {E.~S.}\ \bibnamefont
  {Fradkin}}\ and\ \bibinfo {author} {\bibfnamefont {A.~A.}\ \bibnamefont
  {Tseytlin}},\ }\bibfield  {title} {\bibinfo {title} {{One Loop Beta Function
  in Conformal Supergravities}},\ }\href
  {https://doi.org/10.1016/0550-3213(82)90481-3} {\bibfield  {journal}
  {\bibinfo  {journal} {Nucl. Phys. B}\ }\textbf {\bibinfo {volume} {203}},\
  \bibinfo {pages} {157} (\bibinfo {year} {1982}{\natexlab{a}})}\BibitemShut
  {NoStop}%
\bibitem [{\citenamefont {Fradkin}\ and\ \citenamefont
  {Tseytlin}(1982{\natexlab{b}})}]{Fradkin:1982xc}%
  \BibitemOpen
  \bibfield  {author} {\bibinfo {author} {\bibfnamefont {E.~S.}\ \bibnamefont
  {Fradkin}}\ and\ \bibinfo {author} {\bibfnamefont {A.~A.}\ \bibnamefont
  {Tseytlin}},\ }\bibfield  {title} {\bibinfo {title} {{ASYMPTOTIC FREEDOM IN
  EXTENDED CONFORMAL SUPERGRAVITIES}},\ }\href
  {https://doi.org/10.1016/0370-2693(82)91018-8} {\bibfield  {journal}
  {\bibinfo  {journal} {Phys. Lett. B}\ }\textbf {\bibinfo {volume} {110}},\
  \bibinfo {pages} {117} (\bibinfo {year} {1982}{\natexlab{b}})},\ \bibinfo
  {note} {[Erratum: Phys.Lett.B 126, (1983)]}\BibitemShut {NoStop}%
\bibitem [{\citenamefont {Woodard}(2015)}]{Woodard:2015zca}%
  \BibitemOpen
  \bibfield  {author} {\bibinfo {author} {\bibfnamefont {R.~P.}\ \bibnamefont
  {Woodard}},\ }\bibfield  {title} {\bibinfo {title} {{Ostrogradsky's theorem
  on Hamiltonian instability}},\ }\href
  {https://doi.org/10.4249/scholarpedia.32243} {\bibfield  {journal} {\bibinfo
  {journal} {Scholarpedia}\ }\textbf {\bibinfo {volume} {10}},\ \bibinfo
  {pages} {32243} (\bibinfo {year} {2015})},\ \Eprint
  {https://arxiv.org/abs/1506.02210} {arXiv:1506.02210 [hep-th]} \BibitemShut
  {NoStop}%
\bibitem [{\citenamefont {Langlois}(2019)}]{Langlois:2018dxi}%
  \BibitemOpen
  \bibfield  {author} {\bibinfo {author} {\bibfnamefont {D.}~\bibnamefont
  {Langlois}},\ }\bibfield  {title} {\bibinfo {title} {{Dark energy and
  modified gravity in degenerate higher-order scalar{\textendash}tensor (DHOST)
  theories: A review}},\ }\href {https://doi.org/10.1142/S0218271819420069}
  {\bibfield  {journal} {\bibinfo  {journal} {Int. J. Mod. Phys. D}\ }\textbf
  {\bibinfo {volume} {28}},\ \bibinfo {pages} {1942006} (\bibinfo {year}
  {2019})},\ \Eprint {https://arxiv.org/abs/1811.06271} {arXiv:1811.06271
  [gr-qc]} \BibitemShut {NoStop}%
\bibitem [{\citenamefont {Motohashi}\ \emph {et~al.}(2016)\citenamefont
  {Motohashi}, \citenamefont {Suyama},\ and\ \citenamefont
  {Takahashi}}]{Motohashi:2016prk}%
  \BibitemOpen
  \bibfield  {author} {\bibinfo {author} {\bibfnamefont {H.}~\bibnamefont
  {Motohashi}}, \bibinfo {author} {\bibfnamefont {T.}~\bibnamefont {Suyama}},\
  and\ \bibinfo {author} {\bibfnamefont {K.}~\bibnamefont {Takahashi}},\
  }\bibfield  {title} {\bibinfo {title} {{Fundamental theorem on gauge fixing
  at the action level}},\ }\href {https://doi.org/10.1103/PhysRevD.94.124021}
  {\bibfield  {journal} {\bibinfo  {journal} {Phys. Rev. D}\ }\textbf {\bibinfo
  {volume} {94}},\ \bibinfo {pages} {124021} (\bibinfo {year} {2016})},\
  \Eprint {https://arxiv.org/abs/1608.00071} {arXiv:1608.00071 [gr-qc]}
  \BibitemShut {NoStop}%
\bibitem [{\citenamefont {Myrzakulov}\ \emph {et~al.}(2016)\citenamefont
  {Myrzakulov}, \citenamefont {Sebastiani}, \citenamefont {Vagnozzi},\ and\
  \citenamefont {Zerbini}}]{Myrzakulov:2015kda}%
  \BibitemOpen
  \bibfield  {author} {\bibinfo {author} {\bibfnamefont {R.}~\bibnamefont
  {Myrzakulov}}, \bibinfo {author} {\bibfnamefont {L.}~\bibnamefont
  {Sebastiani}}, \bibinfo {author} {\bibfnamefont {S.}~\bibnamefont
  {Vagnozzi}},\ and\ \bibinfo {author} {\bibfnamefont {S.}~\bibnamefont
  {Zerbini}},\ }\bibfield  {title} {\bibinfo {title} {{Static spherically
  symmetric solutions in mimetic gravity: rotation curves and wormholes}},\
  }\href {https://doi.org/10.1088/0264-9381/33/12/125005} {\bibfield  {journal}
  {\bibinfo  {journal} {Class. Quant. Grav.}\ }\textbf {\bibinfo {volume}
  {33}},\ \bibinfo {pages} {125005} (\bibinfo {year} {2016})},\ \Eprint
  {https://arxiv.org/abs/1510.02284} {arXiv:1510.02284 [gr-qc]} \BibitemShut
  {NoStop}%
\bibitem [{\citenamefont {Gorji}\ \emph {et~al.}(2020)\citenamefont {Gorji},
  \citenamefont {Allahyari}, \citenamefont {Khodadi},\ and\ \citenamefont
  {Firouzjahi}}]{Gorji:2020ten}%
  \BibitemOpen
  \bibfield  {author} {\bibinfo {author} {\bibfnamefont {M.~A.}\ \bibnamefont
  {Gorji}}, \bibinfo {author} {\bibfnamefont {A.}~\bibnamefont {Allahyari}},
  \bibinfo {author} {\bibfnamefont {M.}~\bibnamefont {Khodadi}},\ and\ \bibinfo
  {author} {\bibfnamefont {H.}~\bibnamefont {Firouzjahi}},\ }\bibfield  {title}
  {\bibinfo {title} {{Mimetic black holes}},\ }\href
  {https://doi.org/10.1103/PhysRevD.101.124060} {\bibfield  {journal} {\bibinfo
   {journal} {Phys. Rev. D}\ }\textbf {\bibinfo {volume} {101}},\ \bibinfo
  {pages} {124060} (\bibinfo {year} {2020})},\ \Eprint
  {https://arxiv.org/abs/1912.04636} {arXiv:1912.04636 [gr-qc]} \BibitemShut
  {NoStop}%
\bibitem [{\citenamefont {Dom{\`e}nech}\ \emph {et~al.}(2026)\citenamefont
  {Dom{\`e}nech}, \citenamefont {Ganz},\ and\ \citenamefont
  {Tsabodimos}}]{Domenech:2025gao}%
  \BibitemOpen
  \bibfield  {author} {\bibinfo {author} {\bibfnamefont {G.}~\bibnamefont
  {Dom{\`e}nech}}, \bibinfo {author} {\bibfnamefont {A.}~\bibnamefont {Ganz}},\
  and\ \bibinfo {author} {\bibfnamefont {A.}~\bibnamefont {Tsabodimos}},\
  }\bibfield  {title} {\bibinfo {title} {{On the consistent disformal couplings
  to fermions}},\ }\href {https://doi.org/10.1088/1475-7516/2026/02/045}
  {\bibfield  {journal} {\bibinfo  {journal} {JCAP}\ }\textbf {\bibinfo
  {volume} {02}},\ \bibinfo {pages} {045}},\ \Eprint
  {https://arxiv.org/abs/2510.07419} {arXiv:2510.07419 [hep-th]} \BibitemShut
  {NoStop}%
\bibitem [{\citenamefont {Chamseddine}\ and\ \citenamefont
  {Mukhanov}(2013{\natexlab{b}})}]{Chamseddine_2013}%
  \BibitemOpen
  \bibfield  {author} {\bibinfo {author} {\bibfnamefont {A.~H.}\ \bibnamefont
  {Chamseddine}}\ and\ \bibinfo {author} {\bibfnamefont {V.}~\bibnamefont
  {Mukhanov}},\ }\bibfield  {title} {\bibinfo {title} {Mimetic dark matter},\
  }\bibfield  {journal} {\bibinfo  {journal} {Journal of High Energy Physics}\
  }\textbf {\bibinfo {volume} {2013}},\ \href
  {https://doi.org/10.1007/jhep11(2013)135} {10.1007/jhep11(2013)135} (\bibinfo
  {year} {2013}{\natexlab{b}})\BibitemShut {NoStop}%
\bibitem [{\citenamefont {Ezquiaga}\ and\ \citenamefont
  {Zumalac{\'a}rregui}(2017)}]{Ezquiaga:2017ekz}%
  \BibitemOpen
  \bibfield  {author} {\bibinfo {author} {\bibfnamefont {J.~M.}\ \bibnamefont
  {Ezquiaga}}\ and\ \bibinfo {author} {\bibfnamefont {M.}~\bibnamefont
  {Zumalac{\'a}rregui}},\ }\bibfield  {title} {\bibinfo {title} {{Dark Energy
  After GW170817: Dead Ends and the Road Ahead}},\ }\href
  {https://doi.org/10.1103/PhysRevLett.119.251304} {\bibfield  {journal}
  {\bibinfo  {journal} {Phys. Rev. Lett.}\ }\textbf {\bibinfo {volume} {119}},\
  \bibinfo {pages} {251304} (\bibinfo {year} {2017})},\ \Eprint
  {https://arxiv.org/abs/1710.05901} {arXiv:1710.05901 [astro-ph.CO]}
  \BibitemShut {NoStop}%
\end{thebibliography}%
\end{document}